%% file: main.tex
\documentclass[journal]{IEEEtran}
\usepackage{cite}
\usepackage{amssymb}
\usepackage{amsfonts}
\usepackage{fancyhdr}  %
\usepackage{extarrows}
\usepackage{algorithm}
\usepackage{multirow,tabularx}
\usepackage{mathtools}
\usepackage{xcolor}
\usepackage{bm}
\usepackage{algorithmic}
\usepackage{amsmath}
\usepackage{subfigure}
\usepackage{makecell}
\usepackage{colortbl}
\usepackage{bbm}
\usepackage{enumitem}
\usepackage{acronym}

\allowdisplaybreaks

\ifCLASSINFOpdf
  \usepackage[pdftex]{graphicx}
  \graphicspath{{../pdf/}{../jpeg/}}
  \DeclareGraphicsExtensions{.pdf,.jpeg,.png}
\else
  \usepackage[dvips]{graphicx}
  \graphicspath{{../eps/}}
  \DeclareGraphicsExtensions{.eps}
\fi
\makeatletter
\newenvironment{breakablealgorithm}
  {
   \begin{center}
     \refstepcounter{algorithm}
     \hrule height.8pt depth0pt \kern2pt
     \renewcommand{\caption}[2][\relax]{
       {\raggedright\textbf{\ALG@name~\thealgorithm} ##2\par}%
       \ifx\relax##1\relax 
         \addcontentsline{loa}{algorithm}{\protect\numberline{\thealgorithm}##2}%
       \else 
         \addcontentsline{loa}{algorithm}{\protect\numberline{\thealgorithm}##1}%
       \fi
       \kern2pt\hrule\kern2pt
     }
  }{
     \kern2pt\hrule\relax
   \end{center}
  }
\makeatother

\input{acronyms}

\begin{document}

\title{Extended Target Imaging in Non-Ideal Distributed ISAC Systems: A Hierarchical Off-Grid Approach}

\author{Xin~Tong,~\IEEEmembership{Member,~IEEE,}
        Yu~Ge,~\IEEEmembership{Member,~IEEE,}
        Alireza~Pourafzal,~\IEEEmembership{Member,~IEEE,}
        Dario~Tagliaferri,~\IEEEmembership{Member,~IEEE,}
        Henk~Wymeersch,~\IEEEmembership{Fellow,~IEEE} 
        \vspace{-1cm}

\thanks{Xin~Tong, Alireza~Pourafzal and Henk~Wymeersch are with the Department of Electrical Engineering, Chalmers University of Technology, Gothenburg, Sweden. (e-mails: \{xinto, alireza.pourafzal, henkw\}@chalmers.se)}

\thanks{Yu~Ge is with the Research Laboratory of Electronics, Massachusetts
Institute of Technology, Cambridge, MA, USA. (e-mail: yuge@mit.edu)}

\thanks{Dario~Tagliaferri is with the Department of Electronics, Information and Bioengineering, Politecnico di Milano, 20133, Milan, Italy. (e-mail: dario.tagliaferri@polimi.it)}

}

\maketitle

\begin{abstract}
In future integrated sensing and communication (ISAC) systems, distributed multiple-input multiple-output (MIMO) architectures offer high-resolution imaging capabilities. However, reconstructing real-world extended targets is physically challenging due to continuous anisotropic surface scattering and complex occlusions, especially in the presence of system synchronization errors. To overcome these limitations, this paper proposes a hierarchical two-stage computational imaging framework for extended target reconstruction.
In Phase I, an incoherent coarse imaging process is performed to serve as a reliable initializer for subsequent high-resolution coherent image generation. By performing line-of-sight (LOS) interference suppression and leveraging network-wide topological constraints, timing and phase offsets are calibrated. This enables image construction by back-projection (BP) and the coarse identification of target contours using the probabilistic Hough transform (PHT). In Phase II, a joint target position and orientation off-grid compensation model is established, capturing Lambertian diffuse and specular reflections on continuous facets via a first-order Taylor expansion. To tackle severe occlusion and spatial grid mismatches, we propose the bilinear off-grid approximate message passing (BiOG-AMP) algorithm. BiOG-AMP iteratively updates surface visibility and dictionary matrices, enforcing physical consistency across distributed transceiver views.
Theoretical analysis derives compressed sensing error bounds, Fisher information limits, and linearization error scaling. Simulation results demonstrate that the proposed framework effectively eliminates imaging artifacts and exhibits superior robustness against synchronization offsets and grid mismatches compared to state-of-the-art baselines.
\end{abstract}

\vspace{-0.5cm}
\IEEEpeerreviewmaketitle

\section{Introduction}
\IEEEPARstart{I}{n} future \ac{ISAC} systems, high-precision \ac{RF} imaging is essential for fine-grained environment sensing \cite{JAzhang}. \ac{RF} imaging enables rich geometric situational sensing, which is vital for emerging applications such as autonomous driving, smart factories, and human-robot interaction \cite{Liu}. 
To satisfy resolution demands in these scenarios, distributed \ac{MIMO} architectures serve as a key enabler \cite{Liu, Behdad, TongDistributed}. By deploying spatially separated \acp{Tx} and \acp{Rx}, distributed networks capture multi-perspective scattering responses to significantly enhance imaging resolution.
In practical \ac{ISAC} deployments, \ac{LOS} and \ac{NLOS} paths co-exist. Rather than acting as impairments, \ac{NLOS} multipaths offer additional spatial measurements. Exploiting these signals with multi-view diversity enables the reconstruction of continuous extended target contours \cite{xing}.

Despite these opportunities, transitioning to high-resolution extended target imaging in \ac{D-ISAC} introduces fundamental challenges. The first challenge stems from off-grid mismatch and anisotropic scattering on continuous target surfaces. Unlike point-target imaging with discrete off-grid positions \cite{tong2026nlos}, extended targets feature continuous geometries with unknown positions and surface orientations. Moreover, their scattering is highly anisotropic, making complex scattering coefficients explicitly dependent on observation angle \cite{pegoraro2025hierarchical, pegoraro2026toward}. Consequently, the sensing matrix inherently depends on continuous orientation variables, requiring joint position-orientation mismatch compensation.

The second challenge arises from occlusion effect \cite{XinOcclusion}. In extended target imaging, this effect manifests in two distinct forms: self-occlusion, where the extended target is undetectable due to the directionality of its own surface reflections; and inter-target occlusion, where foreground targets occlude the visibility of background objects. Crucially, because surface visibility dynamically changes the sensing matrix support during surface reconstruction, standard sparse recovery or optimization methods fail to enforce these geometric constraints, inevitably yielding physically inconsistent results such as surface discontinuities, artifacts, or unrealistic penetration.

The third challenge involves synchronization errors and non-convexity in coherent imaging \cite{Dariojsac, han2025over}. Independent \acp{LO} in distributed architectures inevitably introduce \acp{TO} and \acp{PO}. For extended targets, unknown synchronization errors are strongly coupled with geometry, anisotropic scattering, and link visibility. Direct optimization is thus highly unstable, as phase ambiguities easily trap algorithms in physically distorted local solutions.

Motivated by these challenges, we propose a hierarchical two-stage imaging framework with joint position-orientation off-grid compensation for \ac{D-ISAC} systems. To bypass strong non-convexity, Phase I uses robust coarse processing to estimate initial geometry and calibrate synchronization errors. Phase II then performs fine coherent reconstruction with dynamic occlusion constraints and off-grid refinement, achieving high-resolution target shape and scattering profile estimation.

\vspace{-0.3cm}
\subsection{Related Works}
The evolution of environment sensing in \ac{ISAC} systems has been inspired by traditional radar technology, expanding from basic parameter estimation to high-resolution spatial mapping \cite{Liu, ZhangEnabling}. Compared with pure localization, imaging-based sensing provides a richer representation of the environment, including spatial structures, target shapes, and scattering distributions. In distributed architectures, these imaging systems are generally categorized into coherent and incoherent paradigms. Coherent imaging, such as multi-static back-projection and holographic reconstruction, utilizes the phase consistency across spatially separated nodes to achieve high spatial resolution \cite{Zhi2025nearfield, yang2025nearfield}. However, coherent processing is highly sensitive to \acp{TO} and \acp{PO}, requiring strict synchronization \cite{YangPhase, tong2026nlos}. In contrast, incoherent processing \cite{Han2} utilizes phase-insensitive measurements, such as signal amplitudes or power profiles, to circumvent synchronization errors. Although incoherent schemes offer robust optimization with smoother objective functions, they lack phase information, resulting in limited spatial resolution.

Building upon these processing paradigms, recent imaging research has advanced significantly across diverse \ac{ISAC} system architectures. In single-device scenarios, existing works focus on the \ac{ISAC} waveform design for joint communication and radar imaging \cite{zheng2024random}, lightweight linear image formation \cite{Li2021lightweight}, and beam-sweeping or ray-tracing techniques \cite{Bazzi2025isac}. To overcome spatial coverage bottlenecks and mitigate physical occlusion, distributed \ac{D-ISAC} networks have increasingly gained traction, casting imaging as sparse reconstruction problems over pixel grids \cite{XinOcclusion}. These networked paradigms further enable near-field multistatic reconstruction \cite{Zhi2025nearfield} and covariance-based multi-view fusion for extended targets \cite{Gao}.

To improve performance beyond conventional linear methods, computational imaging \cite{Sun, tongx} has emerged as a powerful paradigm for wireless environment sensing. By discretizing the \ac{ROI} into pixels and recovering their scattering coefficients, this approach transforms environmental mapping into a sparse reconstruction problem. Based on compressed sensing theory, various sparse reconstruction algorithms, such as \ac{OMP} \cite{omp} and \ac{GAMP} \cite{Rangan}, have been successfully applied to \ac{OFDM} sensing. Despite its flexibility, computational imaging inherently relies on phase-coherent integration and remains susceptible to synchronization errors. Nevertheless, compared to traditional linear methods, the optimization framework of computational imaging provides a flexible mathematical structure that allows for algorithmic advancements to mitigate these synchronization errors. Existing computational imaging methods can mitigate \acp{PO} via iterative calibration \cite{tong2026nlos}. However, they rely on simplified isotropic point-target models, assuming scattering is independent of target geometry and observation angles.

Despite these advancements, existing computational imaging methodologies are primarily restricted to the isotropic point-target assumption, which simplifies the environment into isolated grid points. However, physical objects in practical scenarios are extended targets with continuous surfaces, specific shapes, and complex geometries. For such targets, physical properties like geometric occlusion must be considered \cite{XinOcclusion}. To address this, several target shape estimation and multi-view profile reconstruction schemes have been developed to trace target boundaries \cite{Maity, Pedraza}. Additionally, advanced off-grid compensation methods have been investigated to handle spatial discretization errors for structured shapes \cite{TongDistributed}. These existing methods typically recover discrete point clusters first and then connect them to form surfaces, rather than treating the object as an extended target directly during the optimization process \cite{xing, zhou}. However, no framework jointly models continuous position-orientation mismatch, anisotropic scattering, and dynamic visibility within coherent reconstruction under distributed synchronization errors.
\vspace{-0.3cm}
\subsection{Contributions}
In this paper, we propose a two-stage computational imaging framework for extended target reconstruction under synchronization offsets in \ac{D-ISAC} systems. Rather than point-target models \cite{Gao, XinOcclusion, tong2026nlos}, we adopt an oriented-facet target model with continuous position and orientation refinement. To overcome severe non-convexity, Phase I deliberately trades resolution for robustness, leveraging incoherent processing and \ac{PHT}) to achieve coarse synchronization and geometric initialization, thereby bypassing phase periodicity traps and making coherent estimation tractable. Phase II then transitions to coherent processing for fine reconstruction. It establishes a joint position-orientation off-grid model alongside a visibility-constrained \ac{BiOG-AMP} algorithm, dynamically updating surface visibility across views to eliminate imaging artifacts.
The main contributions of this paper are summarized as follows:
\begin{itemize}
    \item We propose a hierarchical two-stage imaging framework that transitions from robust incoherent initialization to coherent reconstruction, resolving the non-convex coupling where both the sensing dictionary and visibility pattern depend on unknown surface geometry.

    \item In Phase I, we develop a geometric prior extraction scheme that suppresses strong \ac{LOS} interference and integrates \ac{BP} with \ac{PHT} to initialize the coarse layout and boundaries of extended targets under synchronization impairments.

    \item In Phase II, we propose a joint position and orientation off-grid compensation model that incorporates orientation offsets to capture direction-dependent reflection amplitudes, establishing a physically accurate dictionary for anisotropic continuous surfaces.

    \item We design the \ac{BiOG-AMP} algorithm, which integrates visibility constraints into the sparse recovery process. The algorithm maintains physical consistency by updating occlusion states and off-grid estimation throughout the iterations.
\end{itemize}


\begin{figure}[t]
    \centering
    \includegraphics[width=0.65\linewidth]{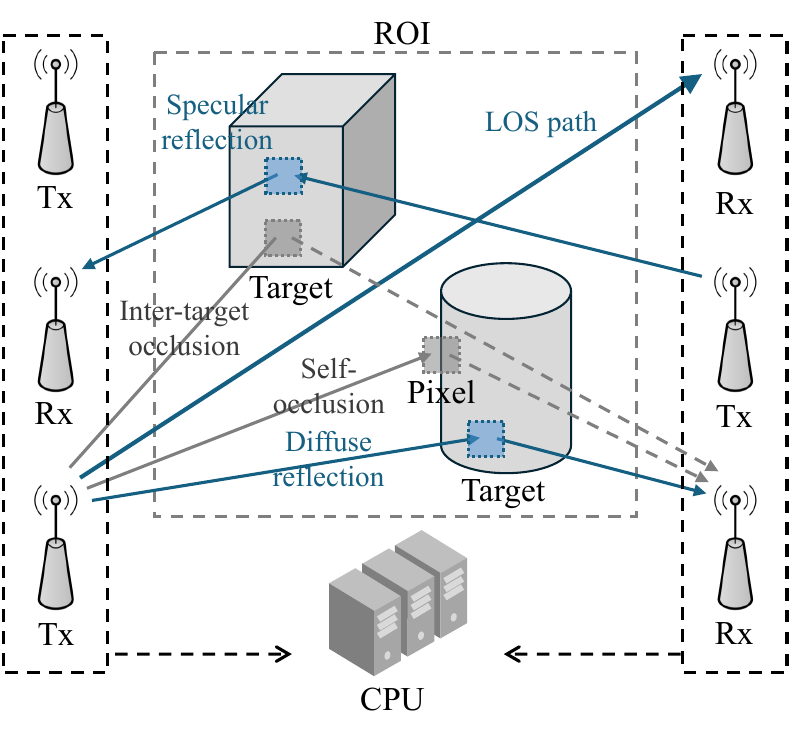}
    \caption{Extended target imaging scenario for \ac{D-ISAC} systems. Blue line: valid LOS and NLOS paths. Gray line: occluded paths.}
    \label{fig:scenario}
\end{figure}

\vspace{-0.3cm}
\section{System Model}
\subsection{Environment Setting and Synchronization Model}
We consider the \ac{D-ISAC} system in Fig.~\ref{fig:scenario}, comprising $N_{\text{Tx}}$ single-antenna \acp{Tx} and $N_{\text{Rx}}$ single-antenna \acp{Rx} deployed at \acp{BS} or distributed as \acp{UE}. All nodes connect to a central \ac{CPU}, forming $M = N_{\text{Tx}} \times N_{\text{Rx}}$ links. For the $m$-th link, the \ac{Tx} and \ac{Rx} positions are denoted by $\bm{p}_{{\rm Tx}, m} \in \mathbb{R}^2$ and $\bm{p}_{{\rm Rx}, m} \in \mathbb{R}^2$.

The system aims to perform high-resolution imaging of multiple extended targets within an \ac{ROI}. The environment includes target-reflected \ac{NLOS} paths and direct \ac{LOS} links, while clutter outside the \ac{ROI} is neglected, as its power is relatively low and can be incorporated into the additive noise. Due to low wave penetration, scattering occurs mainly on target surfaces, leading to inter-target occlusion and orientation-dependent self-occlusion.

Under an \ac{OFDM} framework
, independent \acp{LO} introduce link-dependent \acp{TO} and random carrier \acp{PO}, severely impairing phase coherence. To address this, the \ac{CPU} aggregates \ac{CSI} across $M$ links and $K$ subcarriers for joint processing. Meanwhile, \ac{LOS} links enable \ac{OTA} synchronization to estimate and calibrate phase errors, assisting precision imaging.

\begin{figure*}[!t]
    \begin{equation}
    y_{m,k} = \Big(v_{m}^{\text{LOS}}\alpha_{m}^{\text{LOS}}e^{-j2\pi f_{k}\tau_{m}^{\text{LOS}}} + \sum_{n=1}^{N} v_{m,n}^{\text{NLOS}} x_{n} \alpha_{m}^{\text{NLOS}} (\bm p_{n}, \vartheta_n)e^{-j2\pi f_{k}\tau_{m}^{\text{NLOS}}(\bm p_{n})}\Big)e^{j\theta_{m,k}} + w_{m,k}.\label{eq:ymk}
    \end{equation}
\hrulefill
\vspace{-0.7cm}
\end{figure*}

\vspace{-0.3cm}
\subsection{Computational Imaging Model for Extended Targets}\label{sec:model}
Following traditional computational imaging, the \ac{ROI} of size $L \times W$ is uniformly discretized into $N = (L/\ell) \cdot (W/w)$ pixels of dimensions $\ell \times w$.
The position of the $n$-th pixel is denoted by $\bm p_n$.
Its environmental property is represented by a reflectivity coefficient $x_n \in \mathbb{C}$, where $x_n = 0$ indicates the absence of a target.
To model extended targets, we enhance this framework to account for link-dependent anisotropic scattering caused by target geometry and varying observation angles.
Specifically, each pixel $n$ is characterized by a tuple $(x_n, \vartheta_n)$, where $x_n \in \mathbb{C}$ is the material-dependent base reflectivity and $\vartheta_n$ is the surface orientation angle. Incorporating $\vartheta_n$ renders the effective reflectivity direction-dependent across links.


In large-scale communication scenarios, the continuous target surface is discretized into localized pixels and characterized as electrically large patches rather than infinitesimal scattering points to ensure computational tractability \cite{TongDistributed}. Furthermore, this physical model justifies the integration of specular reflection components alongside diffuse scattering.

Considering the $m$-th link and $k$-th subcarrier, the received \ac{CSI} $y_{m,k}$ is modeled as the superposition of the \ac{LOS} component and the \ac{NLOS} components from the visible pixels of the extended target as shown in \eqref{eq:ymk},
where higher-order multi-hop scattering components are neglected due to significant propagation attenuation and are absorbed into the noise $w_{m,k}$. The amplitude attenuation of \ac{LOS} and \ac{NLOS} propagation are denoted as $\alpha^{\rm LOS}_{m} \in \mathbb{R}$ and $\alpha^{\rm NLOS}_{m}(\bm p_{n}, \vartheta_n) \in \mathbb{R}$.
\begin{equation}
    \alpha^{\rm LOS}_{m} = \frac{\lambda_{c}}{4\pi{\|\bm p_{{\rm Rx},m} - \bm p_{{\rm Tx},m} \|_2 }}, \label{eq:los_path_loss}
\end{equation}
where $\lambda_c$ is the wavelength at the center frequency $f_c$. 
In practical scenarios, the amplitude attenuation of the \ac{NLOS} path $\alpha_{m}^{\rm NLOS}(\bm p_{n}, \vartheta_n)$ depends on the target's reflection mechanism. 

\begin{figure}[t]
    \centering
    \includegraphics[width=0.7\linewidth]{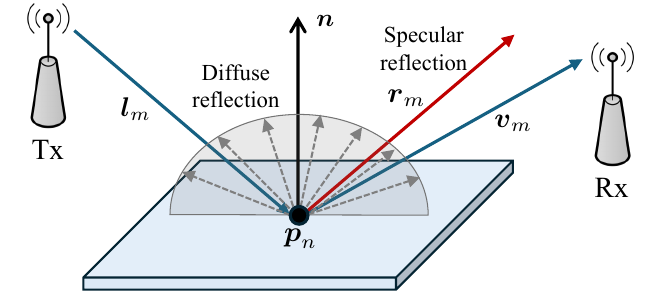}
    \caption{The Lambertian model demonstrates isotropic scattering and the direction of strongest reflection.}
    \label{fig:Lambertian}
\end{figure}

As shown in Fig.~\ref{fig:Lambertian}, to capture extended target anisotropy in RF bands, we adopt a hybrid reflection model inspired by Lambertian scattering model \cite{Degli, Garcia}. Here, a quasi-isotropic diffuse component $k_d$ which, arising from roughness diffraction maintains baseline multi-view observability, while a specular component $k_s$ captures directional energy upon specular alignment. The \ac{NLOS} amplitude attenuation $\alpha_m^{\mathrm{NLOS}}(p_n, \vartheta_n)$ is modeled as
\begin{equation}
    \alpha_{m}^{\rm NLOS}(\bm p_{n}, \vartheta_n) = \alpha_{m}^{\rm diff}(\bm p_{n}, \vartheta_n) + \alpha_{m}^{\rm spec}(\bm p_{n}, \vartheta_n),
\end{equation}
where
\begin{align}
\alpha_{m}^{\rm diff}(\bm p_{n}, \vartheta_n) &= \frac{\sqrt{k_{\rm d} \max(0,\bm n(\vartheta_n) \cdot -\bm l_{m}(\bm p_{n}))}}{L^{\rm d}_{m}(\bm p_{n})}, \\
\alpha_{m}^{\rm spec}(\bm p_{n}, \vartheta_n) &= \frac{\sqrt{k_{\rm s} (\max(0, \bm r_{m}(\bm p_{n}, \vartheta_n) \cdot \bm v_{m}(\bm p_{n})))^{\gamma}}}{L^{\rm s}_{m}(\bm p_{n})},
\end{align}
where $k_{\rm d}$ and $k_{\rm s}$ are the diffuse and specular coefficients and $k_{\rm d} + k_{\rm s} = 1$. The perfect specular reflection vector is calculated as $\bm{r}_m(\bm p_{n}, \vartheta_n) = 2(\bm{n}(\vartheta_n) \cdot \bm{l}_m(\bm p_{n}))\bm{n}(\vartheta_n) - \bm{l}_m(\bm p_{n})$, unit vector $\bm n(\vartheta_n)$ is the pixel normal, and $\gamma$ is the shininess constant which controls the spatial sharpness of the specular reflection lobe. The operator $\max(0, \cdot)$ enforces the local surface orientation condition, ensuring that a pixel contributes to the scattered field only when its reflective surface faces toward both the Tx and Rx. The inter-pixel occlusion is separately addressed during the reconstruction stage via the visibility indicators $v_{m,n}^{\rm NLOS}$, which will be described later. The attenuation of scattering and specular reflection components is 
\begin{align}
L^{\rm d}_m &= 4\pi({\|\bm p_{n} - \bm p_{{\rm Tx},m} \|_2 \times \|\bm p_{n} - \bm p_{{\rm Rx},m} \|_2})/\lambda_{c}, \\
L^{\rm s}_m &= 4\pi({\|\bm p_{n} - \bm p_{{\rm Tx},m} \|_2 + \|\bm p_{n} - \bm p_{{\rm Rx},m} \|_2})/\lambda_{c}.
\end{align}
where all attenuation terms remain dimensionless and unit-consistent.
Indicators $v_{m}^{\rm LOS} \in \mathbb{B}$ and $v_{m,n}^{\rm NLOS} \in \mathbb{B}$ denote \ac{LOS} and \ac{NLOS} visibility indicators, respectively.
$\tau^{\rm LOS}_{m}$ and $\tau^{\rm NLOS}_{m}(\bm p_{n})$ denote propagation delay of \ac{LOS} and \ac{NLOS} paths, 
\begin{equation}
\tau_{m}^{\rm LOS} = \| \bm p_{{\rm Tx},m} - \bm p_{{\rm Rx},m} \|/c,
\end{equation}
\begin{equation}
\tau^{\rm NLOS}_{m}(\bm p_{n}) = \left( \| \bm p_{n} - \bm p_{{\rm Tx},m} \| + \| \bm p_{n} - \bm p_{{\rm Rx},m} \| \right)/c. \label{eq:tNLOS}
\end{equation}
Phase angle term $\theta_{m,k} \sim {\mathcal U}[0,2\pi)$ models the synchronization error for the $m$-th link,
\begin{equation}
\theta_{m,k} = -2\pi f_k \epsilon_{m} + \phi_{m},
\end{equation}
where $\epsilon_{m}$ denotes \ac{TO}, $\phi_{m}$ denotes \ac{PO}.
Physically, these offsets originate from the independent \ac{LO} mismatches at each individual antenna. Let $i$ and $j$ denote the indices of the \ac{Tx} and \ac{Rx} antennas corresponding to the $m$-th link. The link-level synchronization errors can be expressed as the sum of their respective antenna-level errors
\begin{align}
    \epsilon_m &= \epsilon_{{\rm Tx}, i} - \epsilon_{{\rm Rx}, j},\\
    \phi_m &= \phi_{{\rm Tx}, i} - \phi_{{\rm Rx}, j}.
\end{align}
We arbitrarily designate the first \ac{Tx} antenna as the global reference node, forcing its synchronization errors to zero ($\epsilon_{{\rm Tx}, 1} = 0$ and $\phi_{{\rm Tx}, 1} = 0$). 

\begin{figure}[t]
    \centering
    \includegraphics[width=0.8\linewidth]{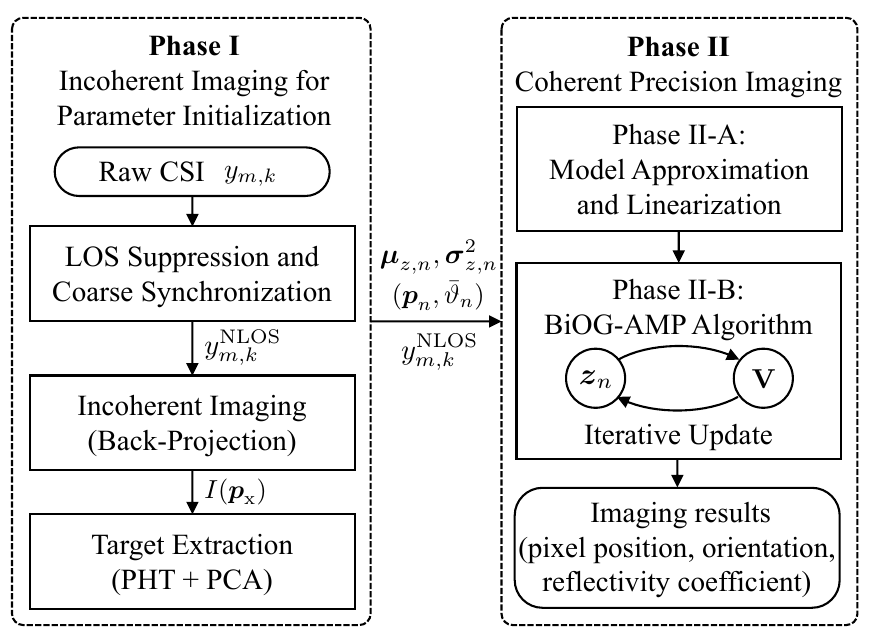}
    \caption{The proposed hierarchical two-stage reconstruction framework.}
    \label{fig:structure}
\end{figure}

To solve the formulated imaging problem under synchronization errors, a hierarchical two-stage reconstruction framework is developed as illustrated in Fig.~\ref{fig:structure}. Taking the multi-link CSI as input, Phase~I performs incoherent global initialization using a phase-insensitive objective function to capture the coarse target profile. Subsequently, Phase~II achieves wavelength-level coherent precision imaging on a fine-grained.

\vspace{-0.3cm}
\section{Phase I: Incoherent Imaging for Parameter Initialization}\label{sec:PhaseI}
Due to phase periodicity, coherent imaging is severely non-convex and prone to local optima. We propose a coarse incoherent initialization strategy, first applying \ac{LOS} suppression and coarse synchronization, followed by incoherent \ac{BP} to identify target regions and orientations.
\vspace{-0.3cm}
\subsection{LOS Suppression and Coarse Synchronization}
In a typical \ac{ISAC} scenario, the \ac{LOS} component often dominates the received signal power, potentially masking the much weaker \ac{NLOS} reflections from targets. To reveal the target-related information, it is crucial to accurately detect and suppress the \ac{LOS} interference. We transform the frequency-domain \ac{CSI} $\bm{y}_m = [y_{m,1}, \dots, y_{m,K}]^\mathsf{T} \in \mathbb{C}^K$ into the delay domain via the \ac{IFFT} to obtain the discrete time-domain impulse response,
\begin{equation}
    \bm h_{m} = \text{IFFT}\{\bm y_{m}\}.
\end{equation}
Exploiting the knowledge of the relative positions of the ISAC nodes in the network, the theoretical delay index corresponding to the \ac{LOS} path is calculated as $n_{\rm t} = \text{round}(\tau_m^{\rm LOS} B)$, where $B$ is the bandwidth. To account for discrete sampling offsets, we search for the peak power within a narrow window centered at $n_{\rm t}$,
\begin{equation}
    P^{\rm LOS}_{m} = \max_{n_{\rm p} \in [n_{\rm t}-\delta_{\rm t}, n_{\rm t}+\delta_{\rm t}]} |h_{m,n_{\rm p}}|^2,
\end{equation}
where $\delta_{\rm t}$ is the searching range on the order of the \ac{CP}. The \ac{LOS} component is declared detected if $P^{\rm LOS}_{m} > \eta_{\rm L} (\alpha_m^{\rm LOS})^2,$ where the theoretical \ac{LOS} path loss $\alpha_m^{\rm LOS}$ is deterministically calculated from the known node positions using \eqref{eq:los_path_loss}, and the discount factor $\eta_{\rm L}$ is set to $0.8$ in this paper to tolerate practical channel noise.

Upon detecting the presence of a dominant \ac{LOS} path, its power significantly overwhelms the weaker \ac{NLOS} target reflections. Consequently, the phase behavior of the received \ac{CSI} $y_{m,k}$ is dictated by the \ac{LOS} propagation delay and synchronization errors, with the \ac{NLOS} components and noise acting merely as small phase noises. Thus, the unwrapped phase $\Psi_{m,k} = \angle y_{m,k}$ of the raw \ac{CSI} can be linearized with respect to the subcarrier frequency $f_k$ as
\begin{equation}
\Psi_{m,k} \approx -2\pi f_k (\tau_m^{\rm LOS} + \epsilon_m) + \phi_m + \omega_{m,k},
\end{equation}
where $\omega_{m,k}$ denotes the residual phase noise caused by the multipath \ac{NLOS} scattering components and additive channel noise.
We perform a linear regression on the observed data set $\{f_k, \Psi_{m,k}\}_{k=1}^{K}$ to determine the slope $\kappa_m$ and the intercept $\beta_m$, that is $\Psi_{m,k} = \kappa_m f_k + \beta_m$.
By aligning the fitting coefficients with the physical model, the coarse synchronization parameters \ac{TO} $\hat{\epsilon}^{\rm LOS}_m$ for the $m$-th link with \ac{LOS} detection is estimated as
\begin{equation}
\hat{\epsilon}^{\rm LOS}_m = -\frac{\kappa_m}{2\pi} - \tau_m^{\rm LOS}.
\end{equation}

Next, we refine the synchronization results by exploiting the topological constraints across the network nodes. Define the global antenna-level \ac{TO} vector as $\bm{e}_{\epsilon} = [\epsilon_{{\rm Tx}, 2}, \dots, \epsilon_{{\rm Tx}, N_{\rm Tx}}, \epsilon_{{\rm Rx}, 1}, \dots, \epsilon_{{\rm Rx}, N_{\rm Rx}}]^{\mathsf T} \in \mathbb{R}^{N_{\rm Tx}+N_{\rm Rx}-1}$, where the first \ac{Tx} antenna is chosen as the global phase/time reference. To achieve phase coherence across all nodes, we solve a linear inverse problem to estimate the antenna-level error vector $\bm e_{\epsilon}$ from the link-level observations. Note that for link pairs without an active \ac{LOS} path, direct coarse estimation is unavailable. Therefore, the least-squares (LS) estimation of $\bm e_{\epsilon}$ is performed using exclusively the subset of valid \ac{LOS}-detected links. Once the global antenna-level errors $\bm e_{\epsilon}$ are recovered, the synchronization offsets for all links can be fully reconstructed and calibrated via the incidence matrix $\mathbf{G}$.

We define an incidence matrix $\mathbf{G} \in \{-1, 0, 1\}^{M \times (N_{Tx}+N_{Rx}-1)}$, where for the $m$-th link, the elements for the \ac{Tx} and \ac{Rx} nodes are set to $+1$ and $-1$, respectively. The estimation is formulated as the following \ac{LS} optimization
\begin{equation}
\hat{\bm e}_{\epsilon} = \arg \min_{\bm e_{\epsilon}} \| \boldsymbol{\epsilon} - \mathbf{G} \bm{e}_{\epsilon} \|_2^2. \label{eq:LS}
\end{equation}
Since the $M$ links are physically formed by only $N_{\rm Tx} + N_{\rm Rx} - 1$ antennas, the true link-level errors must lie within the column space of the incidence matrix $\mathbf{G}$.
Let $\hat{\bm{\epsilon}}^{\rm LOS} = [\hat{\epsilon}^{\rm LOS}_1, \dots, \hat{\epsilon}^{\rm LOS}_M]^{\mathsf T}$ be the vector of initial \ac{TO} estimates. The optimal antenna-level \ac{TO} vector $\hat{\bm{e}}_{\epsilon}$ in \eqref{eq:LS} is obtained by 
\begin{equation}
\hat{\bm{e}}_{\epsilon} = (\mathbf{G}_{\rm sub}^{\mathsf T}\mathbf{G}_{\rm sub})^{-1} \mathbf{G}_{\rm sub}^{\mathsf T} \hat{\bm{\epsilon}}^{\rm LOS}. \label{eq:ee}   
\end{equation}
where $\mathbf{G}_{\rm sub}$ and $\hat{\boldsymbol{\epsilon}}^{\rm LOS}$ consist only of valid links with \ac{LOS} detections, the Tx-Rx LOS graph should be connected, thus enabling inverse operations. Subsequently, the refined link-level \ac{TO} estimates for all links, including those occluded without \ac{LOS}, are reconstructed as $\hat{\boldsymbol{\epsilon}}_{\rm r} = \mathbf{G}\hat{\bm{e}}_{\epsilon}$.

With the globally refined \ac{TO} $\hat{\epsilon}_{r,m}$ establishing a rigid phase slope, we recalculate the optimal \ac{PO} for the valid \ac{LOS} links by extracting the mean phase residual,
\begin{equation}
\hat{\phi}_{m}^{\rm LOS} = \angle \Big( \sum_{k=1}^{K} \exp \big( j \left[ \Psi_{m,k} + 2\pi f_k (\tau_m^{\rm LOS} + \hat{\epsilon}_{{\rm r},m}) \right] \big) \Big).
\end{equation}
For \ac{NLOS} links where direct \ac{PO} observation is impossible due to the lack of a dominant path, we infer their \acp{PO} via the network topology. Specifically, since the antenna-level \ac{PO} vector $\hat{\bm e}_{\phi}$ is recovered from the overdetermined system of all available \ac{LOS} links, the \ac{PO} for any \ac{NLOS}-only link $m$ can be reconstructed as the relative difference between its constituent \ac{Tx} and \ac{Rx} node-level offsets, i.e., $\hat{\phi}_{m} = [\mathbf{G}\hat{\bm e}_{\phi}]_m$. Similar to \eqref{eq:ee}, we obtain $\hat{\bm{e}}_{\phi}$. Note the $2\pi$ phase wrapping ambiguity when applying linear inversion, ensuring that the phase residuals are correctly contained within $[-\pi, \pi]$.
\vspace{-0.3cm}
\subsection{Incoherent Imaging via Space-Time Matched Filter}
Using these synchronized estimates, the target-related \ac{NLOS} \ac{CSI} component $y_{m,k}^{\rm NLOS}$ is extracted as
\begin{equation}
y_{m,k}^{\rm NLOS} = 
\begin{cases}
y_{m,k} \exp\big(j (2\pi f_k \hat{\epsilon}_{r,m} - \hat{\phi}_{r,m})\big) \\ \quad\quad - \alpha_{m}^{\rm LOS} \exp\big(-j2\pi f_k \tau_m^{\rm LOS}\big), v_m^{\rm LOS} = 1, \\
y_{m,k} \exp\big(j (2\pi f_k \hat{\epsilon}_{r,m} - \hat{\phi}_{r,m})\big),  v_m^{\rm LOS} = 0,
\end{cases}
\end{equation}
where $v_m^{\rm LOS} \in \{0, 1\}$ indicates the presence of an active \ac{LOS} path. Specifically, for links with detected \ac{LOS} components, the signal is synchronized and the dominant geometric \ac{LOS} reference is subtracted; for \ac{LOS}-occluded links, the synchronized signal serves directly as the calibrated \ac{NLOS} output.

Following this \ac{LOS} suppression and hardware synchronization, the purified \ac{NLOS} signals across all subcarriers are transformed into the delay domain to extract the \ac{PDP}, which subsequently serves as the input for the \ac{BP} imaging in Phase I to capture the coarse geometric layout of the extended targets.

We transform the frequency-domain $y_{m,k}^{\rm NLOS}$ into the delay domain to extract the \ac{PDP}. For the $m$-th link, the incoherent power envelope at a theoretical delay $\tau$ is given by
\begin{equation}
    P_{m}(\tau) = \Big| \sum_{k=1}^K y_{m,k}^{\rm NLOS} e^{j 2\pi f_k \tau} \Big|^2.
\end{equation}
We first establish a coarse spatial grid covering the sensing environment and construct the \ac{BP} map. The energy intensity at a coarse grid point ${\bm p}_{{\rm x}}$ is accumulated over all subcarriers and links
\begin{equation}
I({\bm p}_{{\rm x}}) = \sum_{m=1}^M P_m(\hat{\tau}^{\rm NLOS}_m({\bm p}_{{\rm x}})),
\end{equation}
where $\hat{\tau}^{\rm NLOS}_m$ is the theoretical bi-static propagation delay over link $m$ for a hypothetical target as location $\bm p_{\rm x}$ which can be calculated by \eqref{eq:tNLOS}.
\vspace{-0.3cm}
\subsection{Target Extraction}\label{sec:TE}
This section extracts target geometry from the coarse Phase I image to initialize the high-resolution imaging in Phase II. Differing from conventional point-target assumptions where scatterers are modeled as isolated pixels, extended targets physically occupy multiple contiguous pixels with smooth boundaries. Motivated by this physical characteristic, and following the approach in \cite{xing}, we apply \ac{PHT} to identify the principal linear boundaries from the coarse image. We first apply a threshold $I_{\rm t}$ to the \ac{BP} image $I$ to generate a binary mask $\mathcal{B}$, where $\mathcal{B}(\bm{p}) = 1$ if the intensity exceeds $I_{\rm t}$. To identify line segments within the noisy mask, we employ the \ac{PHT}.
The \ac{SHT} maps each pixel $(x, y)$ in the image space to a curve in the $(\rho_{\rm l}, \theta_{\rm l})$ parameter space according to
\begin{equation}
\rho_{\rm l} = x \cos \theta_{\rm l} + y \sin \theta_{\rm l},
\end{equation}
where $\theta_{\rm l} \in [0, \pi)$ is the angle of the normal to the line and $\rho_{\rm l}$ is the distance from the origin. While the \ac{SHT} considers all pixels, the \ac{PHT} randomly samples a subset of points to detect discrete line segments rather than infinite lines. The output of this step is a set of initial segments $\mathcal{L}_{\rm r} = \{L_1, \dots, L_{N_{\rm l}}\}$.

Due to the coarse resolution of Phase I, a single physical target may appear as several parallel or fragmented segments after \ac{PHT}. We merge these fragments into unified clusters $\mathcal{C}_q$ based on their spatial and angular proximity. Each cluster $\mathcal{C}_q$ represents a reconstructed unified physical extended target. Specifically, two segments $L_i$ and $L_j$ are assigned to the same cluster $\mathcal{C}_{q}$ if they satisfy the following consistency criteria
\begin{equation}
\mathcal{D}(L_i, L_j) = \mathbb{I} \left( |\theta_i - \theta_j| < \epsilon_{\theta} \right) \cdot \mathbb{I} \left( \|\bm{m}_i - \bm{m}_j\|_2 < \epsilon_{d} \right),
\end{equation}
where $\bm{m}_i$ is the midpoint of segment $L_i$, while $\epsilon_{\theta}$ and $\epsilon_{d}$ represent the angular and Euclidean distance tolerances.

For each cluster $\mathcal{C}_q$, let $\bm{p}_{a,L}, \bm{p}_{b,L} \in \mathbb{R}^2$ denote the two endpoints of a segment $L \in \mathcal{C}_q$. We aggregate all endpoints into a point set $\mathbf{P}_q = \bigcup_{L \in \mathcal{C}_q} \{\bm{p}_{a,L}, \bm{p}_{b,L}\}$. The spatial extent and orientation of target $q$ are estimated via \ac{PCA} on the covariance matrix
\begin{equation}
\bm{\Sigma}_q = \frac{1}{|\mathbf{P}_q|} \sum_{\bm{p} \in \mathbf{P}_q} (\bm{p} - \bar{\bm{p}}_q)(\bm{p} - \bar{\bm{p}}_q)^{\mathsf T},
\end{equation}
where $\bar{\bm{p}}_q = \frac{1}{|\mathbf{P}_q|} \sum_{\bm{p} \in \mathbf{P}_q} \bm{p}$ is the spatial centroid. 
Let $\bm{u}_{1} = [u_{1,x}, u_{1,y}]^{\mathsf T}$ be the primary eigenvector of $\bm{\Sigma}_q$. The refined target endpoints $\bm{p}_{\text{start},q}$ and $\bm{p}_{\text{end},q}$ are obtained by projecting $\mathbf{P}_q$ onto $\bm{u}_1$:
\begin{align}
\bm{p}_{\text{start},q} &= \bar{\bm{p}}_q + \min_{\bm{p} \in \mathbf{P}_q} \left( (\bm{p} - \bar{\bm{p}}_q)^{\mathsf T} \bm{u}_1 \right) \bm{u}_1, \\
\bm{p}_{\text{end},q} &= \bar{\bm{p}}_q + \max_{\bm{p} \in \mathbf{P}_q} \left( (\bm{p} - \bar{\bm{p}}_q)^{\mathsf T} \bm{u}_1 \right) \bm{u}_1.
\end{align}
The refined surface normal is $\bar{\vartheta}_q = \text{atan2}(u_{1,y}, u_{1,x}) + {\pi}/{2}$.

The definite output of this stage for each target $q$ is the line segment tuple $\{\bm{p}_{\text{start},q}, \bm{p}_{\text{end},q}, \bar{\vartheta}_q\}$. These outputs directly serve as the geometric prior for Phase II to initialize pixel positions $\bar{\bm p}_n$ along the line segment and set their surface normals as $\bar{\vartheta}_n = \bar{\vartheta}_q$.
\vspace{-0.3cm}
\section{Phase II-A: Coherent Precision Imaging Model}
In Phase II, we refine the Phase I initialization via coherent imaging within the identified \ac{ROI}. Accounting for environmental continuity and anisotropic scattering, our model jointly refines reflectivity, off-grid offsets, and link-pixel occlusion status.

\newcounter{mytempeqncnt}
\begin{figure*}[!t]
\normalsize
\setcounter{mytempeqncnt}{\value{equation}}
\setcounter{equation}{33}
\begin{align}
    \frac{\partial a_{m,k}}{\partial \vartheta_n} =  \frac{1}{\ell} \int_{-\ell/2}^{\ell/2} \Big[ \Big( \frac{\partial \alpha_{m}^{\rm diff}(\bar{\bm{p}}_{n,\rho}, \bar{\vartheta}_n)}{\partial \vartheta_n} + \frac{\partial \alpha_{m}^{\rm spec}(\bar{\bm{p}}_{n,\rho}, \bar{\vartheta}_n)}{\partial \vartheta_n} \Big) + j \frac{2\pi \rho}{\lambda_k} \alpha_m^{\rm NLOS}(\bar{\bm{p}}_{n,\rho}, \bar{\vartheta}_n) \left( \bm{u}_{{\rm Tx},m}(\rho) + \bm{u}_{{\rm Rx},m}(\rho) \right) \cdot \bm{n} \Big] \nonumber \\ \times e^{-j2\pi f_{k}\tau_{m}^{\rm NLOS}(\bar{\bm{p}}_{n,\rho})} d\rho.\label{eq:at}
\end{align}
\setcounter{equation}{\value{mytempeqncnt}}
\hrulefill
\vspace{-0.6cm}
\end{figure*}

\begin{figure}[t]
    \centering
    \includegraphics[width=0.5\linewidth]{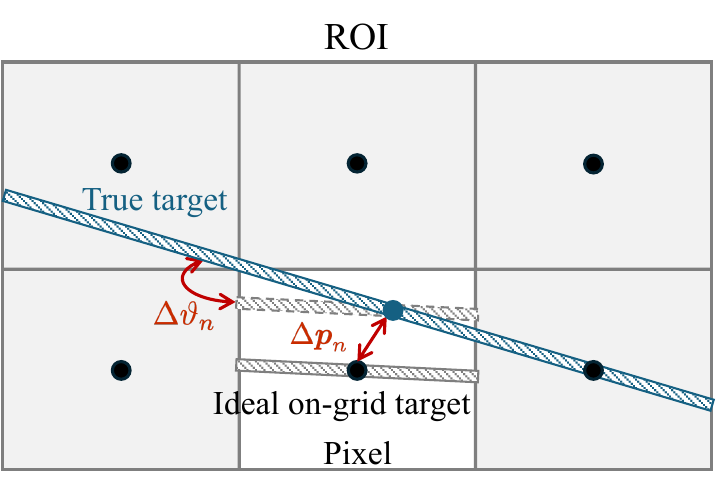}
    \caption{The offset between the real target and the ideal target represented by pixels, including the offset of position and orientation.}
    \label{fig:off-grid}
\end{figure}

\vspace{-0.3cm}
\subsection{Pixel Model for Extended Target}
For the $m$-th link and $k$-th subcarrier, the observed \ac{NLOS} \ac{CSI} $y^{\rm NLOS}_{m,k}$ in \eqref{eq:ymk} can be reformulated as
\begin{equation}
y^{\rm NLOS}_{m,k} = \sum_{n=1}^N v_{m,n}^{\rm NLOS} x_n a_{m,k}(\bm p_n, \vartheta_n) + w_{m,k},
\end{equation}
where the propagation kernel $a_{m,k}(\bm p_n, \vartheta_n) = \alpha_{m}^{\rm NLOS}(\bm p_{n}, \vartheta_n)e^{\Phi_{m,k}(\bm p_{n})}$ combines phase $\Phi_{m,k}(\bm p_{n}) = -j2\pi f_{k}\tau_{m}^{\rm NLOS}(\bm p_{n})$ and amplitude $\alpha_{m}^{\rm NLOS}(\bm p_n, \vartheta_n)$, and $w_{m,k}$ is the additive white Gaussian noise with variance $\sigma^2_{\rm n}$.

To accommodate the continuous nature of extended targets, we model each discrete pixel as a spatial facet rather than a point, thereby reformulating the propagation kernel $a_{m,k}(\bm p_n, \vartheta_n)$. Each pixel $n$ in Phase II is modeled as a continuous linear facet of length $\ell$ along its target profile rather than a point. Let $\rho \in [-\ell/2, \ell/2]$ denote the local coordinate along the segment representing the $n$-th pixel facet. Under the orientation $\vartheta_n$, the spatial location of a micro-element on this facet can be calculated as
\begin{equation}
    {\bm{p}}_{n, \rho} = {\bm{p}}_n + \rho \bm{t}({\vartheta}_n),
\end{equation}
where ${\bm{p}}_n$ is the grid center of the $n$-th pixel, and $\bm{t}({\vartheta}_n) = [-\sin{\vartheta}_n, \cos{\vartheta}_n]^{\mathsf T}$ represents the unit tangent vector of the facet . 
Consequently, the propagation kernel $a_{m,k}({\bm{p}}_n, {\vartheta}_n)$ is no longer evaluated at a single point, but is defined as the spatial line integral across the entire facet length
\begin{align}
    a_{m,k}&({\bm{p}}_n, {\vartheta}_n) = \nonumber\\ & \frac{1}{\ell} \int_{-\ell/2}^{\ell/2} \alpha_{m}^{\rm NLOS}({\bm{p}}_{n, \rho},{\vartheta}_n) e^{-j2\pi f_{k}\tau_{m}^{\rm NLOS}({\bm{p}}_{n, \rho})} d\rho, \label{eq:integral}
\end{align}
where $\alpha_{m}^{\rm NLOS}({\bm{p}}_{n, \rho}, {\vartheta}_n)$ and $\tau_{m}^{\rm NLOS}({\bm{p}}_{n, \rho})$ represent the Lambertian amplitude attenuation and the propagation delay of the micro-element at ${\bm{p}}_{n,\rho}$, respectively.
\vspace{-0.3cm}
\subsection{First-Order Taylor Approximation and Linearization}\label{sec:Taylor}
In practical scenarios, the true target may deviate from grids, while also taking into account the coarse estimation error in Phase I. As illustrated in Fig.~\ref{fig:off-grid}, we define $\bar{\bm{p}}_n$ as the reference pixel grid center and $\bar{\vartheta}_n$ as the coarse surface normal initialized by Phase I. The true spatial position and surface orientation of pixel $n$ are represented as $\bm{p}_n = \bar{\bm{p}}_n + \Delta \bm{p}_n$ and ${\vartheta}_n = \bar{\vartheta}_n + \Delta \vartheta_n$, where $\Delta \bm{p}_n$ and $\Delta \vartheta_n$ denote the unknown off-grid spatial and orientation offsets, respectively.
Note that the theoretical propagation kernel $a_{m,k}(\bm p_n, \vartheta_n)$ becomes unknown because it includes an unknown offset. Therefore, we need to approximate this model in this section.

The off-grid offset $\Delta \bm{p}_n$ mainly induces phase rotations, while the orientation offset $\Delta \vartheta_n$ modifies the reflection angle without affecting distance. Applying a first-order Taylor expansion with respect to position and orientation, the \ac{NLOS} propagation gain is approximated as
\begin{align}
a_{m,k}(\bm{p}_n, \vartheta_n) \approx &\; a_{m,k}(\bar{\bm{p}}_n, \bar{\vartheta}_n) + \left(\nabla_{\bm{p}} a_{m,k}\right)^{\mathsf T} \Delta \bm{p}_n \nonumber\\&+ \frac{\partial a_{m,k}}{\partial \vartheta_n} \Delta \vartheta_n.\label{eq:Taylor}
\end{align}
Since $\nabla_{\bm{p}} \alpha_{m}^{\rm NLOS} \ll \alpha_{m}^{\rm NLOS} \nabla_{\bm{p}} \Phi_{m,k}(\bm p_{n})$ at high carrier frequencies, we omit the derivative with respect to position $\bm{p}_n$ of the amplitude attenuation. Let $\bm{u}_{{\rm Tx},m}$ and $\bm{u}_{{\rm Rx},m}$ be the unit vectors pointing from the \ac{Tx} and \ac{Rx} nodes toward the $n$-th pixel center $\bar{\bm{p}}_n$, respectively. The partial derivatives are derived as
\begin{align}
\frac{\partial a_{m, k}}{\partial x_n} \approx & \frac{1}{\ell} \int_{-\ell/2}^{\ell/2} -j \frac{2\pi}{\lambda_{k}} \left( u_{{\rm Tx},m}^{(x)}(\rho) + u_{{\rm Rx},m}^{(x)}(\rho) \right) \nonumber \\
&\alpha_{m}^{\rm NLOS}(\bar{\bm{p}}_{n,\rho}, \bar{\vartheta}_n) e^{-j2\pi f_{k}\tau_{m}^{\rm NLOS}(\bar{\bm{p}}_{n,\rho})} d\rho, \label{eq:ax} \\ 
\frac{\partial a_{m, k}}{\partial y_n} \approx & \frac{1}{\ell} \int_{-\ell/2}^{\ell/2} -j \frac{2\pi}{\lambda_{k}} \left( u_{{\rm Tx},m}^{(y)}(\rho) + u_{{\rm Rx},m}^{(y)}(\rho) \right) \nonumber \\ 
&\alpha_{m}^{\rm NLOS}(\bar{\bm{p}}_{n,\rho}, \bar{\vartheta}_n) e^{-j2\pi f_{k}\tau_{m}^{\rm NLOS}(\bar{\bm{p}}_{n,\rho})} d\rho.\label{eq:ay}
\end{align}
When there is no self-occlusion, let the surface normal vector be $\bm{n} = [\cos\bar{\vartheta}_n, \sin\bar{\vartheta}_n]^{\mathsf T}$, and its orthogonal tangent vector be $\bm{n}' = [-\sin\bar{\vartheta}_n, \cos\bar{\vartheta}_n]^{\mathsf T}$. Rotating the facet about its center $\bar{\bm{p}}_n$ causes the geometric derivative $\frac{\partial \bar{\bm{p}}_{n,\rho}}{\partial \vartheta_n} = -\rho \bm{n}$ for the micro-element at offset $\rho$. Consequently, taking the derivative with respect to $\vartheta_n$ yields both amplitude variations and rotation-induced phase shifts inside the integral, We obtain \eqref{eq:at}. The detailed derivations for the diffuse and specular reflection gradients ${\partial \alpha_{m}^{\rm diff}}/{\partial \vartheta_n}$ and ${\partial \alpha_{m}^{\rm spec}}/{\partial \vartheta_n}$ are deferred to Appendix~\ref{app:reflection_gradients}.
\setcounter{equation}{\value{equation}+1}

To transform this linearized model into a structured sparse recovery problem, we define a state vector for the $n$-th pixel: $\bm{z}_n = [x_n, s_{x,n}, s_{y,n}, s_{\vartheta,n}]^{\mathsf T}$, where $s_{x,n} = x_n \Delta x_n$, $s_{y,n} = x_n \Delta y_n$, and $s_{\vartheta,n} = x_n \Delta \vartheta_n$ are the auxiliary offset variables. For the $m$-th link across all $K$ subcarriers, the \ac{NLOS} \ac{CSI} vector is
\begin{equation}
\bm{y}^{\rm NLOS}_m = \sum_{n=1}^N v^{\rm NLOS}_{m,n} \mathbf
{A}_{m,n} \bm{z}_n + \bm{w}_m,
\end{equation}
where $\bm{a}_{m,n} = [a_{m,1}(\bm{p}_n, \vartheta_n), \dots, a_{m,K}(\bm{p}_n, \vartheta_n)]^{\mathsf{T}} \in \mathbb{C}^K$ collects the channel coefficients across all $K$ subcarriers for the $n$-th pixel, $\mathbf{A}_{m,n} = \left[ \bm{a}_{m,n}, \frac{\partial \bm{a}_{m,n}}{\partial x}, \frac{\partial \bm{a}_{m,n}}{\partial y}, \frac{\partial \bm{a}_{m,n}}{\partial \vartheta} \right] \in \mathbb{C}^{K \times 4}$ is the link-specific dictionary block, and $\bm{w}_m$ is the Gaussian white noise vector. Aggregating all $M$ links, the global imaging model is written as
\begin{equation}
\bm{y}^{\rm NLOS} = \big( (\mathbf{V} \otimes \bm{1}_{K \times 4}) \circ \mathbf{A} \big) \bm{z} + \bm{w},
\end{equation}
where $\bm{y}^{\rm NLOS} \in \mathbb{C}^{MK}$ is the concatenated observation vector, $\mathbf{V} \in \mathbb{B}^{M \times N}$ is the unknown link-pixel occlusion matrix containing the element $v^{\rm NLOS}_{m,n}$, $\mathbf{A} \in \mathbb{C}^{MK \times 4N}$ is the concatenated dictionary matrix, and $\bm{z} = [\bm{z}_1^{\mathsf T}, \dots, \bm{z}_N^{\mathsf T}]^{\mathsf T} \in \mathbb{C}^{4N}$ is the state vector to be recovered.
\vspace{-0.4cm}
\subsection{Problem Formulation}
The joint estimation is formulated as a maximum\textit{ a posteriori} probability problem
\begin{align}
(\hat{\bm{z}}, \hat{\mathbf{V}}) = & \arg \max_{\bm{z}, \mathbf{V}} \Pr( \bm{z}, \mathbf{V} |\bm{y}^{\rm NLOS}; \mathbf{A}(\bm{z})) \label{eq:op1}\\
\text{s.t.} \quad & v_{m,n} = \mathcal{G}_{m,n}(\bm{z}), \quad \forall m, n,
\end{align} 
where $\mathcal{G}_{m,n}(\cdot)$ represents the geometric visibility constraint and the parameter dictionary $\mathbf{A}(\bm{z})$ depends on the target geometry $\bm{z}$. We consider the occlusion into two physical phenomena: self-occlusion (determined by the surface normal) and mutual-occlusion (determined by occlusion from other targets). Therefore, $\mathcal{G}_{m,n}(\cdot)$ is expressed as
\begin{align}\label{eq:G_func}
    \mathcal{G}_{m,n}(\bm{z}) = \;& \mathbb{I} \left( \bm{u}_{{\rm Tx},m}^{\mathsf T} \bm{n}(\vartheta_n) > 0 \right) \cdot \mathbb{I} \left( \bm{u}_{{\rm Rx},m}^{\mathsf T} \bm{n}(\vartheta_n) > 0 \right) \nonumber \\ &\cdot \left( 1 - \mathcal{S}_{m,n} \right),
\end{align}
where $\bm{n}(\vartheta_n)$ is the surface normal vector of $n$-th pixel. The self-occlusion term strictly forces the pixel to be invisible if its reflective surface faces away from either the \ac{Tx} or the \ac{Rx}. The term $\mathcal{S}_{m,n} \in \mathbb{B}$ is a binary shadowing function that evaluates to $1$ if any target intersects the propagation paths $\bm p_{{\rm Tx},m} \to \bm p_n$ or $\bm p_n \to \bm p_{{\rm Rx},m}$.

\begin{figure}[t]
    \centering
    \includegraphics[width=0.9\linewidth]{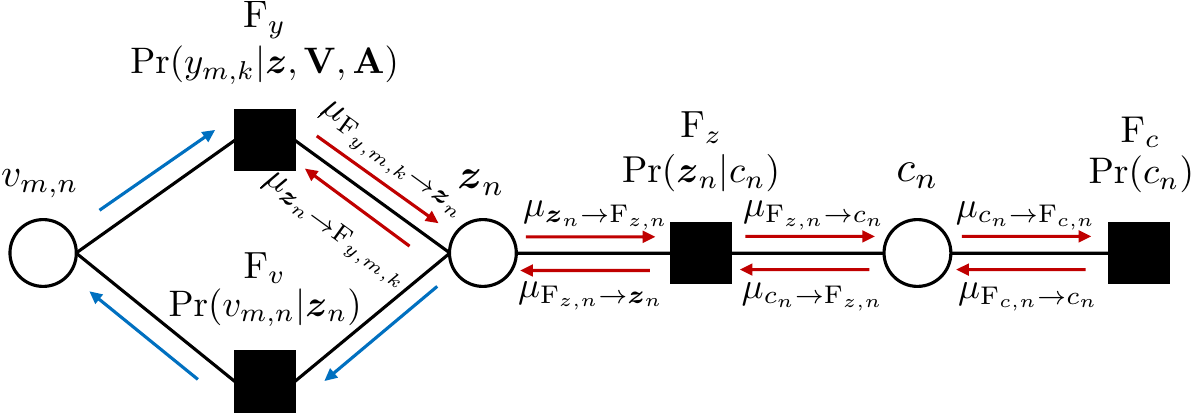}
    \caption{The factor graph of the BiOG-AMP algorithm, where squares represent FNs and circles represent VNs.}
    \label{fig:fg}
\end{figure}

As shown in Fig.~\ref{fig:fg}, to solve this problem efficiently via the proposed \ac{BiOG-AMP}, we design a factor graph based on the decomposition result of the joint posterior probability,
\begin{equation}\label{eq:fg}
\Pr(\bm{z}, \bm{c}, \mathbf{V} | \bm{y}, \mathbf{A}) \propto \Pr(\bm{y} | \bm{z}, \mathbf{V}, \mathbf{A}) \Pr(\mathbf{V} | \bm{z}) \Pr(\bm{z} | \bm{c}) \Pr(\bm{c}),
\end{equation}
where $\bm c \in \mathbb{B}^N$ is an activation index indicating that non-zero elements in $\bm x$, $\bm s_x$, $\bm s_y$, and $\bm s_\vartheta$ have the same index. Note that for clarity, the \ac{NLOS} superscript of $\bm{y}^{\rm NLOS}$ is omitted. We use a factor graph \cite{Kschischang} to represent the posterior probability in \eqref{eq:fg}. In Fig.~\ref{fig:fg}, circles represent \acp{VN} $\bm{z}_n$, $v_n$ and $c_n$, and squares represent \acp{FN} $\Pr(y_{m,k} | \bm{z}, \mathbf{V}, \mathbf{A})$, $\Pr(\bm{z}_n|c_n)$, $\Pr(v_{m,n} | \bm{z}_n)$, and $\Pr(c_n)$, whose names are abbreviated as ${\rm F}_{y,m,k}$, ${\rm F}_{z,n}$, ${\rm F}_{v,m,n}$, and ${\rm F}_{c,n}$, respectively. The lines connecting the nodes indicate that these variables are included in the functions of the \acp{FN}. The specific functions of \acp{FN} are defined as follows:
\begin{itemize}
    \item FN ${\rm F}_{y,m,k}$: The likelihood factor for the $m$-th link under Gaussian noise, yielding $\Pr(y_{m,k} | \bm{z}, \mathbf{V}, \mathbf{A}) = \mathcal{CN} \left( \sum_{n=1}^{N} v_{m,n} \bm{a}_{m,k,n} \bm{z}_n, \sigma_{\rm n}^2 \mathbf{I} \right)$.

    \item FN ${\rm F}_{v,m,n}$: The structural prior node enforcing the physical geometry, defined as $\Pr(v_{m,n} | \bm{z}) = \delta(v_{m,n} - \mathcal{G}(\bm{z}))$ which indicate that visibility is determined by the currently estimated pixel position and orientation.

    \item FN ${\rm F}_{z,n}$: The conditional prior enforcing joint sparsity among the physical state components in $\bm{z}_n$,
        \begin{equation}
        \Pr(\bm{z}_n|c_n) = \left\{
        \begin{aligned}
            &\mathcal{CN}(\bm{\mu}_{z,n}, \bm{\sigma}^2_{z,n}), \quad c_n = 1, \\
            &\delta(\bm{z}_n), \quad c_n = 0,
        \end{aligned}
        \right.
    \end{equation}
    where the variable $c_n$ controls whether component variables ${x}_{n}$, $s_{x,n}$, $s_{y,n}$, and $s_{\vartheta,n}$ in $\bm{z}_n$ are activated together, with means $\bm{\mu}_{z,n} = [\mu_{x,n}, \mu_{s_x,n}, \mu_{s_y,n}, \mu_{s_\vartheta,n}]$ and variances $\bm{\sigma}^2_{z,n} = [\sigma^2_{x,n}, \sigma^2_{s_x,n}, \sigma^2_{s_y,n}, \sigma^2_{s_\vartheta,n}]$ which are set based on the result of Section~\ref{sec:TE}. Specifically, if the pixel range contains the line detected in Phase I, the average position offset is set to the distance from the pixel center to the line, and the average direction is set to the normal direction of the line. Note that in the algorithm, we treat ${\bm z}_{n}$ as a complex value and take the real part in the final estimation. This can absorb common phase offsets, thereby enhancing robustness.

    \item FN ${\rm F}_{c,n}$: Define the sparse prior distribution of targets in the environment as $\Pr(c_n) = \eta^{c_n} (1-\eta)^{1-c_n}$, where $\eta$ is the prior sparsity probability.
\end{itemize}
\vspace{-0.3cm}
\section{Phase II-B: BiOG-AMP Algorithm}\label{sec:BiOG-AMP}
To jointly estimate the state vector $\bm{z}$ and the occlusion matrix $\mathbf{V}$, we propose the \ac{BiOG-AMP} algorithm. Unlike conventional bilinear recovery methods that treat the multiplicative mask $\mathbf{V}$ as unconstrained random variable, our approach couples the occlusion status with the physical geometry of the extended targets to ensure stable convergence and physical consistency.
\vspace{-0.3cm}
\subsection{Approximate Message Passing}\label{sec:AMP}
In this section, we derive the messages iteratively exchanged between \acp{VN} and \acp{FN} based on the sum-product algorithm \cite{Kschischang}. Following the standard belief propagation and \ac{AMP} framework \cite{Rangan}. Specifically, by neglecting higher-order infinitesimal terms and applying the \ac{CLT} to the aggregated inputs at factor nodes, the incoming messages are accurately approximated as Gaussian distributions. Note that the message update rules for $x_n$, $s_{x,n}$, $s_{y,n}$, and $s_{\vartheta,n}$ share an equivalent mathematical structure and follow similar Gaussian derivation steps.

It is important to note that this subsection focuses on the messages associated with the state vector $\bm{z}$ (represented by the red arrows in Fig.~\ref{fig:fg}), while the visibility matrix $\mathbf{V}$ (the blue arrows) is treated as a fixed prior within this stage. The overall framework operates via an alternating AMP and geometric-visibility update scheme. This reduces the oscillations in the cyclic factor graph, thus stabilizing convergence. Furthermore, this helps to traverse the likelihood surface more smoothly, preventing hard decisions from causing the algorithm to get trapped in suboptimal local minima.

\subsubsection{From FN ${\rm F}_{y}$ to VN $\bm z_n$}
The mean and variance of the message $\mu_{{\rm F}_{y,m,k}\rightarrow z_{n,i}}^{(t)}(z_{n,i})$ in $t$-th iteration are updated as
\begin{equation}
    \hat z_{m,k \rightarrow z_{n,i}}^{(t)} = \frac{y_{m,k} - Z_{m,k \rightarrow z_{n,i}}^{(t)}}{a_{m,k,n}^{(i)}},
    \label{eq:z_update_mean}
\end{equation}
\begin{equation}
    v_{m,k \rightarrow z_{n,i}}^{(t)} = \frac{\sigma_{\rm w}^2 + V_{m,k \rightarrow z_{n,i}}^{(t)}}{|a_{m,k,n}^{(i)}|^2},
    \label{eq:z_update_var}
\end{equation}
where $a_{m,k,n}^{(i)}$ denotes the linear coefficient or partial derivative for the $i$-th state variable at the $n$-th grid. The detailed derivation of \eqref{eq:z_update_mean} and \eqref{eq:z_update_var} is provided in Appendix~\ref{ap:A}. Note that since the update structures for all four components in $\bm z_n$ are mathematically equivalent under the first-order Taylor expansion, we utilize $x_n$ as the representative variable in the derivation.

\subsubsection{From VN $\bm z_n$ to FN ${\rm F}_z$}
As derived in Appendix~\ref{ap:B}, the mean and variance of the message $\mu_{z_{n,i} \rightarrow {\rm F}_{z,n}}^{(t)}(z_{n,i})$ for the $i$-th component of the state vector $\bm z_n$ are updated by combining the incoming Gaussian messages from all links and subcarriers:
\begin{align}
\hat{z}_{n,i \rightarrow {\rm F}_{z}}^{(t)} &= v_{n,i \rightarrow {\rm F}_z}^{(t)} \Big( \sum_{m=1}^{M} \sum_{k=1}^{K} \frac{\hat z_{m,k \rightarrow z_{n,i}}^{(t)}}{v_{m,k \rightarrow z_{n,i}}^{(t)}} \Big), \label{eq:z_vn_to_fn_mean} \\
v_{n,i \rightarrow {\rm F}_z}^{(t)} &= \Big( \sum_{m=1}^{M} \sum_{k=1}^{K} \frac{1}{v_{m,k \rightarrow z_{n,i}}^{(t)}} \Big)^{-1}, \label{eq:z_vn_to_fn_var}
\end{align}
Since the derivations for $x_n, s_{x,n}, s_{y,n}$, and $s_{\vartheta,n}$ are similar, we provide the detailed derivation for $x_n$ in the Appendix~\ref{ap:B}.

\subsubsection{From FN ${\rm F}_z$ to VN $c_n$}
The message $\mu^{(t)}_{{\rm F}_{z,n} \rightarrow {c_n}}(c_n)$ at the $t$-th iteration is calculated by marginalizing over all continuous components of the state vector $\bm{z}_n$,
\begin{equation}
\mu^{(t)}_{{\rm F}_{z,n} \rightarrow {c_n}}(c_n) \propto Z_{x,n}^{(t)}(c_n) Z_{s_{x,n}}^{(t)}(c_n) Z_{s_{y,n}}^{(t)}(c_n) Z_{s_{\vartheta,n}}^{(t)}(c_n),
\label{eq:mfzntcn}
\end{equation}
where $Z_{z_{n,i}}^{(t)}(c_n) = \int \Pr(z_{n,i}|c_n) \mu_{z_{n,i}\rightarrow {\rm F}_{z,n}}^{(t)}(z_{n,i}) {\rm d}z_{n,i}$ for each component $i \in \{x, s_x, s_y, s_\vartheta\}$. As derived in Appendix~\ref{ap:C}, the terms $Z_{x,n}^{(t)}(c_n)$ are given by:
\begin{align}
Z_{x,n}^{(t)}(0) &= \mathcal{CN}(0 ; \hat{x}_{n\rightarrow {\rm F}_z}^{(t)}, v_{x_n\rightarrow {\rm F}_z}^{(t)}), \label{eq:zxnt0_final} \\
Z_{x,n}^{(t)}(1) &= \mathcal{CN}(\mu_{x,n} ; \hat{x}_{n\rightarrow {\rm F}_z}^{(t)}, \sigma_{x,n}^2 + v_{x_n\rightarrow {\rm F}_z}^{(t)}),\label{eq:zxnt1_final}
\end{align}
where $\mu_{x,n}$ and $\sigma_{x,n}^2$ are the prior mean and variance of the reflectivity when the pixel is active. The derivations for the offset variables $s_{x,n}$, $s_{y,n}$, and $s_{\vartheta,n}$ are similar and will not be repeated here.

\subsubsection{From FN ${\rm F}_c$ to VN $c_n$}
This message is unrelated to iteration, $\mu_{{\rm F}_{c,n} \rightarrow c_n}(c_n) = \Pr(c_n)$.

\subsubsection{From VN $c_n$ to FN ${\rm F}_z$}
This message is unrelated to iteration, $\mu_{c_n \rightarrow {\rm F}_{z,n}}(c_n) \propto \mu_{{\rm F}_{c,n} \rightarrow c_n}(c_n)$.

\subsubsection{From FN ${\rm F}_z$ to VN $z_n$}
The message $\mu_{{\rm F}_{z,n}\rightarrow z_{n,i}}^{(t+1)}(z_{n,i})$ represents the prior information refined by the sparsity constraint and the other components of the state vector. As derived in Appendix~\ref{ap:D}, the message passed to the \ac{VN} $x_n$ is a Bernoulli-Gaussian distribution,
\begin{equation}
\mu_{{\rm F}_{z,n}\rightarrow x_{n}}^{(t+1)}(x_{n}) \propto k_{n,0}^{(t+1)} \delta(x_n) + k_{n,1}^{(t+1)} \mathcal{CN}(x_{n};\mu_{x,n},\sigma_{x,n}^{2}),
\label{eq:fzntxn_final}
\end{equation}
where the weights $k_{n,0}^{(t+1)}$ and $k_{n,1}^{(t+1)}$ are defined as
\begin{equation}
k_{n,0}^{(t+1)} \propto (1-\eta) \prod_{j \in \{s_x, s_y, s_\vartheta\}} Z_{z_{n,j}}^{(t)}(0),
\label{eq:kn0_final}
\end{equation}
\begin{equation}
k_{n,1}^{(t+1)} \propto \eta \prod_{j \in \{s_x, s_y, s_\vartheta\}} Z_{z_{n,j}}^{(t)}(1),
\label{eq:kn1_final}
\end{equation}
where $Z_{z_{n,j}}^{(t)}(c_n)$ are defined in \eqref{eq:zxnt0_final} and \eqref{eq:zxnt1_final}. The derivations for the offset variables $s_{x,n}$, $s_{y,n}$, and $s_{\vartheta,n}$ are similar and will not be repeated here.

\subsubsection{From VN $z_n$ to FN ${\rm F}_{y}$}
As derived in Appendix~\ref{ap:E}, the mean and variance of the message are given by
\begin{equation}
\hat{z}_{n,i \rightarrow {\rm F}_{y,m,k}}^{(t+1)} = P_{n,i,1 \rightarrow m,k}^{(t+1)} \hat{z}_{{\rm prod}, n,i,m,k}^{(t+1)},
\label{eq:zntfymk_mean}
\end{equation}
\begin{align}
v_{n,i \rightarrow {\rm F}_{y,m,k}}^{(t+1)} &= P_{n,i,1 \rightarrow m,k}^{(t+1)} v_{{\rm prod}, n,i,m,k}^{(t+1)}  \nonumber \\
&+ \left( P_{n,i,1 \rightarrow m,k}^{(t+1)} - (P_{n,i,1 \rightarrow m,k}^{(t+1)})^2 \right) |\hat{z}_{{\rm prod}, n,i,m,k}^{(t+1)}|^2,
\label{eq:zntfymk_var}
\end{align}
where the intermediate variables including the parameters and normalized weights are detailed in Appendix~\ref{ap:E}.
\vspace{-0.3cm}
\subsection{Estimation and Iterative Execution}
\subsubsection{Estimation for Active Pixels}
At the $t$-th iteration, for each grid point $n$, the posterior probability of the sparsity variable $c_n$ being active ($c_n=1$) or inactive ($c_n=0$) is calculated by combining the prior message $\mu_{{\rm F}_{c,n} \to c_n}$ and the message $\mu_{{\rm F}_{z,n} \to c_n}^{(t)}$,
\begin{equation}
\Pr(c_n=c|\bm{y},\mathbf{A}) \propto \eta^c (1-\eta)^{1-c} \prod_{i \in \{x, s_x, s_y, s_\vartheta\}} Z_{z_{n,i}}^{(t)}(c), \quad 
\label{eq:pcn_combined}
\end{equation}
where $c \in \{0,1\}$, $Z_{z_{n,i}}^{(t)}(c_n)$ are defined in \eqref{eq:zxnt0_final} and \eqref{eq:zxnt1_final} for each component of the state vector $\bm{z}_n$. Based on this, we calculate the marginal belief $\pi_n^{(t)} = {\Pr(c_n=1|\bm{y},\mathbf{A})}/{\sum_{c \in \{0,1\}} \Pr(c_n=c|\bm{y},\mathbf{A})}$ and determine active pixels as $\mathcal{A}^{(t)} = \{n \mid \pi_n^{(t)} > \eta_c \}$, where $\eta_c$ is typically set to $0.5$.

\subsubsection{Estimation for the State Vector}
For pixels identified as inactive ($n \notin \mathcal{A}^{(t)}$), the estimates are set to zero, i.e., $\hat{x}_n^{(t)} = 0$, $\hat{s}_{x,n}^{(t)} = 0$, $\hat{s}_{y,n}^{(t)} = 0$, and $\hat{s}_{\vartheta,n}^{(t)} = 0$. For the active pixels ($n \in \mathcal{A}^{(t)}$), the posterior distribution for each component $z_{n,i}$ is calculated as
\begin{align}
&\Pr(z_{n,i} | \bm{y}, \mathbf{A}, c_n=1) \nonumber \\ &\propto \mu_{{\rm F}_{z,n} \to z_{n,i}}(z_{n,i} | c_n=1) \prod_{m,k} \mu_{{\rm F}_{y,m,k} \to z_{n,i}}^{(t)}(z_{n,i}) \nonumber \\
&\propto \mathcal{CN}(z_{n,i}; \mu_{z,n,i}, \sigma_{z,n,i}^2) \mathcal{CN}(z_{n,i}; \hat{z}_{n,i \to {\rm F}_z}^{(t)}, v_{n,i \to {\rm F}_z}^{(t)}),
\end{align}
The resulting posterior mean $\hat{z}_{n,i}^{(t)}$ and variance $v_{n,i}^{(t)}$ are calculated as
\begin{equation}
\hat{z}_{n,i}^{(t)} = v_{n,i}^{(t)} \Big( \frac{\mu_{z,n,i}}{\sigma_{z,n,i}^2} + \frac{\hat{z}_{n,i \to F_z}^{(t)}}{v_{n,i \to F_z}^{(t)}} \Big),
\label{eq:zn_est}
\end{equation}
\begin{equation}
v_{n,i}^{(t)} = \Big( \frac{1}{\sigma_{z,n,i}^2} + \frac{1}{v_{n,i \to F_z}^{(t)}} \Big)^{-1}.
\label{eq:vn_est}
\end{equation}
These refined estimates $\hat{\bm{z}}_n^{(t)}$ are then used to update the dictionary and occlusion matrices for the subsequent iteration.

\subsubsection{Occlusion and Dictionary Matrix Update}
A key feature of the proposed algorithm is that the dictionary matrices and the occlusion state are not static but are dynamically refined throughout the iterations. The visibility $v_{m,n}$ of each pixel $n$ relative to the $m$-th link is updated based on the current belief of the scenario geometry. As indicated by the blue arrows in Fig.~\ref{fig:fg}, the estimated state vector $\hat{\bm{z}}_n^{(t)} = [\hat{x}_n, \hat{s}_{x,n}, \hat{s}_{y,n}, \hat{s}_{\vartheta,n}]^{\mathsf T}$ are fed back to the visibility \ac{FN} ${\rm F}_{v,m,n}$. 

To minimize the first-order Taylor approximation error in \eqref{eq:Taylor}, the vectors $\bm{a}^{(t+1)}$, $ \frac{\partial \bm{a}^{(t+1)}}{\partial x}$, $ \frac{\partial \mathbf{a}^{(t+1)}}{\partial y}$, and $ \frac{\partial \bm{a}^{(t+1)}}{\partial \vartheta}$ are recalculated. By re-evaluating the partial derivatives using the latest offset estimates $\{\hat{s}_{x,n}^{(t)}, \hat{s}_{y,n}^{(t)}, \hat{s}_{\vartheta,n}^{(t)}\}$, the dictionary entries are continuously aligned with the true physical shape of the target, ensuring accurate reconstruction in non-linear scattering scenarios. The occlusion state is updated as $v_{m,n}^{(t+1)} = \mathcal{G}_{m,n}(\hat{\bm{z}}^{(t)})$. When $v_{m,n}^{(t+1)} = 0$, the edge between the corresponding nodes is dropped, applying visibility masking ${\bf A}^{(t+1)} \leftarrow ({\bf V}^{(t+1)} \otimes \mathbf{1}_{K \times 4}) \odot {\bf A}^{(t+1)}$ in the next message passing iteration. This feedback masks blocked paths in the observation factors $F_{y,m,k}$, enabling robust imaging under occlusion effect.

\subsubsection{Iterative Execution}
The message passing rules derived in Section~\ref{sec:AMP} define an iterative algorithm. We summarize the proposed \ac{BiOG-AMP} algorithm in Algorithm~\ref{alg:BiOG-AMP}. 
In each iteration, the primary computational cost arises from matrix-vector multiplications in Steps 3 and 4, requiring $\mathcal{O}(MNK)$ operations. Posterior statistics, partition functions, and weight updates (Steps 5 and 6) involve element-wise operations with complexity $\mathcal{O}(N)$ or $\mathcal{O}(M)$. Infrequent outer-loop updates to the occlusion matrix $\mathbf{V}$ incur negligible cost. Thus, the overall imaging complexity is $\mathcal{O}(MNK)$, scaling linearly with subcarriers and pixels. This offers superior efficiency over matrix-inversion methods like MMSE with $\mathcal{O}((4N)^3)$ complexity given four state components per pixel.


\begin{breakablealgorithm}
\small
  \caption{\vspace{-0.0cm}The proposed BiOG-AMP Algorithm}
  \label{alg:BiOG-AMP}
  \begin{algorithmic}[1]
  \REQUIRE
  Frequencies $f_{k}$; positions $\bm p_{\rm Tx}$, $\bm p_{\rm Rx}$, and $\bm p$; observation $\bm y$; prior parameters $\{\mu_{i,n}, \sigma^2_{i,n}\}$ for $i \in \{x, s_x, s_y, s_{\vartheta}\}$.
  \STATE
  \textbf{Initialization}:  According to the imaging result of Phase I, set $\hat{z}_{n\rightarrow {\rm F}_{y,m,k}}^{(0)} = \mu_{i,n}$, $v_{n\rightarrow {\rm F}_{y,m,k}}^{(0)} = \sigma_{i,n}^2$ for all $i$ and initialize occlusion $\mathbf{V}^{(0)}$. Initialize ${\mathbf{A}}^{(0)}$ according to Section~\ref{sec:model} and Section~\ref{sec:Taylor}. 

  \WHILE {$t<t_{\rm max}$ and $\|\hat{\bm x}^{(t)} - \hat{\bm x}^{(t-1)}\| / N > \varepsilon$, where $\varepsilon$ is a given error tolerance value}
  \STATE
    \textit{Update Messages from FN ${\rm F}_{y} \to$ VN $\bm z_{n}$}:
    \begin{itemize}[leftmargin=*, topsep=0pt, itemsep=0pt, parsep=0pt, partopsep=0pt]
        \item Calculate $Z_{m,k \rightarrow n,i}^{(t)}$ and $V_{m,k \rightarrow n,i}^{(t)}$ for $i \in \{x, s_x, s_y, s_{\vartheta}\}$ following \eqref{eq:zmtxn} and \eqref{eq:vvmtxn}.
        \item Calculate $\hat z_{m,k \rightarrow z_{n,i}}^{(t)}$ and $v_{m,k \rightarrow z_{n,i}}^{(t)}$ following \eqref{eq:z_update_mean} and \eqref{eq:z_update_var}.
    \end{itemize}
  \STATE
    \textit{Update Messages From VN $\bm z_n \to$ FN ${\rm F}_{z}$}:
    \begin{itemize}[leftmargin=*, topsep=0pt, itemsep=0pt, parsep=0pt, partopsep=0pt]
        \item Calculate $\hat{z}_{n,i \to {\rm F}_z}^{(t)}$ and $v_{n,i \to {\rm F}_z}^{(t)}$ for all $i$ following \eqref{eq:z_vn_to_fn_mean} and \eqref{eq:z_vn_to_fn_var}.
    \end{itemize}
  \STATE
    \textit{Update Messages From FN ${\rm F}_{z} \to$ VN $\bm z_n \to$ FN ${\rm F}_{y}$}:
    \begin{itemize}[leftmargin=*, topsep=0pt, itemsep=0pt, parsep=0pt, partopsep=0pt]
        \item Calculate $Z_{z_{n,i}}^{(t)}(c_n)$ for $c_n \in \{0,1\}$ and all components $i$ following \eqref{eq:zxnt0_final} and \eqref{eq:zxnt1_final}.
        \item  Calculate cavity parameters $\hat{z}_{{\rm cav}}^{(t)}$ and $v_{{\rm cav}}^{(t)}$ following \eqref{eq:cavz} and \eqref{eq:cavv}.
        \item Calculate product parameters $\hat{z}_{{\rm prod}}^{(t+1)}$ and $v_{{\rm prod}}^{(t+1)}$ following \eqref{eq:prodz} and \eqref{eq:prodv}.
        \item Calculate weights $k_{n,c}^{(t+1)}$, $W_{n,i,c}^{(t+1)}$ following \eqref{eq:kn0_final}, \eqref{eq:kn1_final}, \eqref{eq:Wni0}, and \eqref{eq:Wni1}. Update normalized probability $P_{n,i,1 \to m,k}^{(t+1)}$ .
        \item Calculate $\hat{z}_{n,i \to m,k}^{(t+1)}$ and $v_{n,i \to m,k}^{(t+1)}$ following \eqref{eq:zntfymk_mean} and \eqref{eq:zntfymk_var}.
    \end{itemize}
  \STATE
    \textit{Estimation and Update}:
    \begin{itemize}[leftmargin=*, topsep=0pt, itemsep=0pt, parsep=0pt, partopsep=0pt]
        \item Estimate active pixels $\mathcal{A}^{(t)}$.
        \item For $n \notin \mathcal{A}^{(t)}$: Set $\hat{\bm z}_n^{(t)} = \bm 0$. Keep gradients $\mathbf{A}^{(t+1)}$ unchanged.
        \item For $n \in \mathcal{A}^{(t)}$: Calculate  $\hat{\bm z}_n^{(t)}$ and $v_n^{(t)}$ following \eqref{eq:zn_est} and \eqref{eq:vn_est}. Calculate normalized shifts $\Delta \hat{x}_n^{(t)} = \hat{s}_{x,n}^{(t)}/\hat{x}_n^{(t)}$, $\Delta \hat{y}_n^{(t)} = \hat{s}_{y,n}^{(t)}/\hat{x}_n^{(t)}$, and $\Delta \hat{\vartheta}_n^{(t)} = \hat{s}_{\vartheta,n}^{(t)}/\hat{x}_n^{(t)}$. Update dictionary $\mathbf{A}^{(t+1)}$ and occlusion $\mathbf{V}^{(t+1)}$.
    \end{itemize}
  \STATE
  Set $t = t + 1$. In each iteration, execute steps~3-5. Every $t'$-th iteration, also execute step~6.
  \ENDWHILE
  \ENSURE
  Estimates $\hat{x}_n$, $\Delta \hat{x}_n$, $\Delta \hat{y}_n$, and $\Delta \hat{\vartheta}_n$.
  \end{algorithmic}
  \end{breakablealgorithm}

\section{Theoretical Performance Analysis}
In this section, we go beyond traditional computational imaging techniques that rely on isotropic point scatterer models, and analyze the theoretical impact of extended targets on system performance. 
\vspace{-0.3cm}
\subsection{Ideal Imaging Performance}\label{sec:perA}
We first analyze the ideal baseline performance by assuming perfect synchronization and grid alignment ($\mathbf{\Phi} = \mathbf{I}$, $\Delta \bm p = \mathbf{0}$, $\Delta \bm \vartheta = \mathbf{0}$). Unlike traditional point-target scattering, extended targets introduce severe self-occlusion and inter-target occlusion. This geometrically distributed network results in incomplete observation, where each target pixel is only illuminated by a portion of the links.

Although the visibility constraints and non-isotropic pixels break the \ac{i.i.d.} properties of the measurement matrix, compressed sensing guarantees stable sparse recovery provided that the effective number of unoccluded and well-aligned measurements satisfies the scaling law $K \sum_{m=1}^{M} v_{m,n} \geq C s \log(N/s)$
where $s = |\mathcal{A}|$ is the sparsity level and $C$ is a constant governed by the \ac{RIP} of the structured dictionary \cite{tong2026nlos}.
Under this condition, the expectation of the reconstruction error can be bounded as
\begin{equation}
{\rm E}\left[\|\hat{{\bm x}}-{\bm x}\|_2^2\right] \propto \frac{\sigma_{\rm n}^2 \eta \log N}{K \sum_{n \in \mathcal{A}} \sum_{m=1}^{M} v_{m,n} |\alpha_{m}(p_n, \vartheta_n)|^2}. 
\end{equation}
Reconstruction accuracy is no longer a simple linear function of the total number of links $M$. Specifically, even if $M$ is large, the estimation error may increase if the target's orientation $\vartheta_n$ causes severe self-occlusion or unfavorable specular reflection angles.
\vspace{-0.3cm}
\subsection{Impact of Extended Geometry on Sensing Performance}\label{sec:perB}
We now analyze the impact of extended target geometry on sensing performance based on the \ac{FI}. The sensing measurement vector associated with the $n$-th target pixel is $\bm a_n = \big[ v_{1,n} \alpha_{1,n} \bm e_{1,n}^{\mathsf T}, \dots, v_{M,n} \alpha_{M,n} \bm e_{M,n}^{\mathsf T} \big]^{\mathsf T} \in \mathbb{C}^{MK \times 1}$, where $\alpha_{m,n} \triangleq \alpha_{m}(p_n, \vartheta_n)$ denotes the reflection gain and $\bm e_{m,n} \in \mathbb{C}^{K \times 1}$ is the subcarrier phase progression vector.

The equivalent \ac{FI} for recovering the $n$-th pixel reflectivity $x_n$ corresponds to the $n$-th diagonal element of the target reflectivity \ac{FIM} $\mathbf J_{\bm x} = \frac{2}{\sigma_{\rm n}^2} \Re (\mathbf A^{\mathsf H} \mathbf A)$, expressed as
\begin{equation}
    [\mathbf J_{\bm x}]_{n,n} = \frac{2}{\sigma_{\rm n}^2} \|\bm a_n\|_2^2 = \frac{2 K}{\sigma_{\rm n}^2} \sum_{m=1}^{M} v_{m,n} |\alpha_{m,n}|^2, \label{eq:J_nn}
\end{equation}
with the estimation error bound satisfying $\mathrm{CRB}(x_n) \geq [\mathbf J_{\bm x}^{-1}]_{n,n}$. Directly from the mathematical structure of \eqref{eq:J_nn}, the impacts of geometric factors are characterized as follows:

\begin{itemize}
    \item Since Lambertian scattering provides a positive, quasi-isotropic diffuse reflection gain $\alpha_{m,n} \approx \alpha_{m,n}^{\text{diff}} > 0$ across visible angles, every unoccluded link ($v_{m,n}=1$) contributes a non-zero positive term to \eqref{eq:J_nn}. Consequently, $[\mathbf J_{\bm x}]_{n,n}$ scales monotonically with the link number $M$, ensuring non-zero diagonal entries and preventing numerical ill-conditioning in $\mathbf J_{\bm x}$.

    \item Specular reflection exhibits narrow angular support with high amplitude ($\alpha_{m,n}^{\text{spec}} \gg \alpha_{m,n}^{\text{diff}}$). When a link satisfies specular alignment, $|\alpha_{m,n}|^2$ increases substantially, injecting a dominant term into the summation in \eqref{eq:J_nn}. This causes a sharp local increase in $[\mathbf J_{\bm x}]_{n,n}$ and a corresponding reduction in $\mathrm{CRB}(x_n)$, yielding enhanced local estimation accuracy.

    \item Structural occlusion acts as a binary mask $v_{m,n} \in \{0, 1\}$. Setting $v_{m,n} = 0$ zeroes out the $m$-th component in \eqref{eq:J_nn}, monotonically reducing the local \ac{FI}. In the case of complete occlusion ($v_{m,n} = 0, \forall m$), we have $[\mathbf J_{\bm x}]_{n,n} = 0$, creating geometric sensing blind zones.
\end{itemize}
\vspace{-0.3cm}
\subsection{Model Approximation Error}\label{sec:perC}

To accommodate continuous target positions and surface orientations on discrete grids, first-order Taylor expansions yield the aggregate linearization error $\varepsilon_{m,k}^{\text{off}} = \varepsilon_{m,k}^{\text{pos}} + \varepsilon_{m,k}^{\text{ori}}$. For spatial position mismatch $\Delta \bm p_n$, the variance of the second-order Taylor remainder under uniform pixel displacement scales as \cite{tong2026nlos},
\begin{equation}
{\rm E}\big[|\varepsilon_{m,k}^{\text{pos}}|^2\big] \propto \left(\frac{2\pi}{\lambda_c}\right)^{4} d_{\mathrm{pixel}}^{4}.
\end{equation}
For extended targets modeled as continuous linear facets of length $d_{\text{pixel}}$, an orientation offset $\Delta \vartheta_n$ induces a local displacement $\Delta \bm r(\rho) \approx \rho \bm d_n \Delta \vartheta_n$ along coordinate $\rho \in [-d_{\text{pixel}}/2, d_{\text{pixel}}/2]$. Integrating this phase perturbation across the facet and evaluating its second-order remainder yields
\begin{equation}
{\rm E}\big[|\varepsilon_{m,k}^{\text{ori}}|^2\big] \propto \left(\frac{2\pi}{\lambda_c}\right)^4 d_{\text{pixel}}^4 {\rm E}\big[|\Delta \vartheta_n|^4\big].
\end{equation}
Consequently, the orientation error inherits the fundamental $\lambda_c^{-4} d_{\text{pixel}}^4$ scaling from conventional spatial grid mismatch, while introducing a quantization penalty proportional to the fourth moment of the orientation offset ${\rm E}\big[\vert{}\Delta \vartheta_n\vert{}^4\big]$.
\vspace{-0.3cm}
\section{Numerical Results}
In this section, we evaluate the performance of the proposed imaging framework through numerical simulations. The numerical results demonstrate the effectiveness of the proposed approach in imaging performance for extended targets. Furthermore, we compare the proposed method with two representative benchmark schemes: \ac{GAMP} \cite{Rangan} and \ac{OG-AMP} \cite{tong2026nlos}. It is worth noting that, compared with the classical \ac{OMP} algorithm, \ac{AMP}-based algorithms exhibit stronger robustness and superior reconstruction performance in compressed sensing problems with non-strictly Gaussian measurement matrices. Consequently, \ac{AMP}-based methods serve as the primary baseline for our performance evaluations \cite{TongDistributed}.

\begin{figure*}[h]
  \centering
  \begin{minipage}[t]{0.22\textwidth}
      \centering
      \subfigure[Phase I imaging results under perfect synchronization, including the extracted target. (${\rm CD} = 0.3848~{\rm m}^2$)]{
      \includegraphics[width=0.9\textwidth]{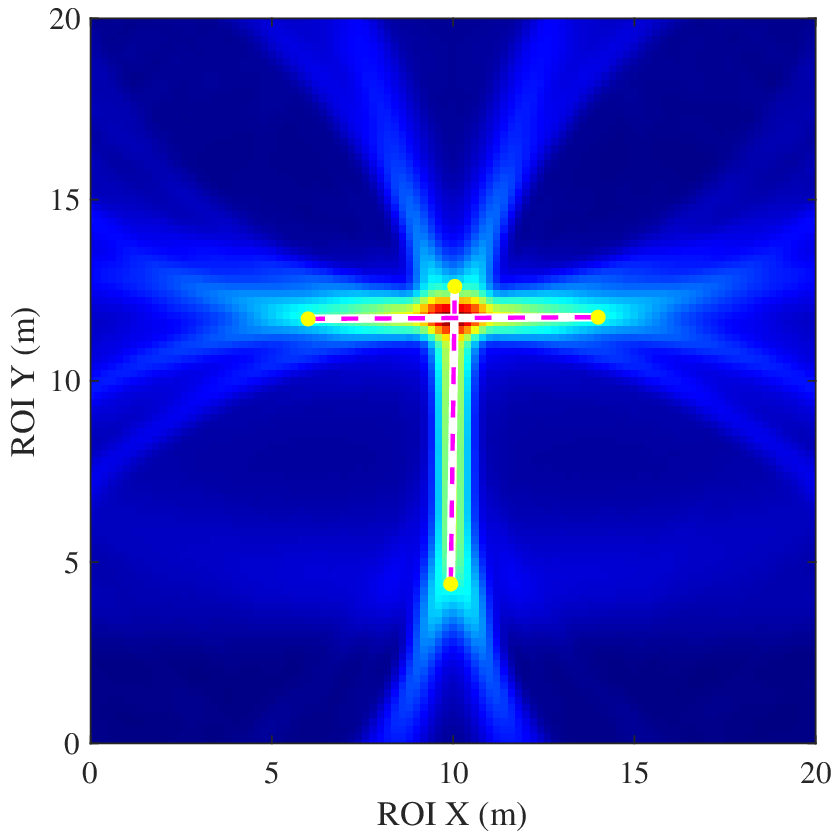}}
  \end{minipage}
  \begin{minipage}[t]{0.01\textwidth}\end{minipage}
  \begin{minipage}[t]{0.22\textwidth}
    \centering
    \subfigure[Unsynchronized Phase I imaging results, including the extracted target. ($\rm TO = 5~ns,~CD = 1.2811~m^2$)]{
    \includegraphics[width=0.9\textwidth]{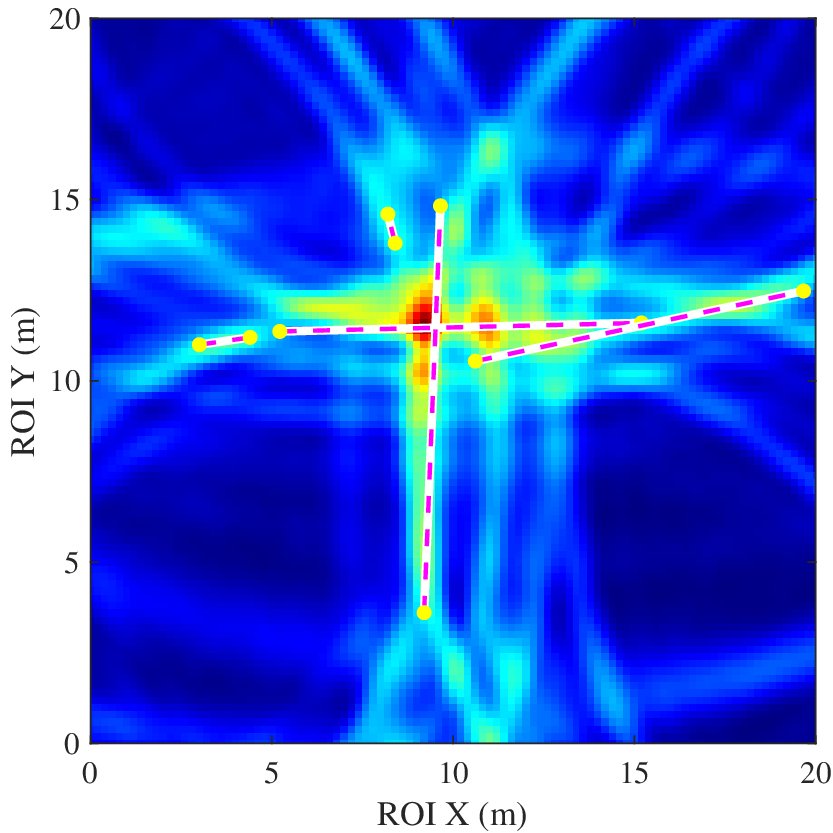}}
\end{minipage}
\begin{minipage}[t]{0.01\textwidth}\end{minipage}
\begin{minipage}[t]{0.22\textwidth}
    \centering
    \subfigure[Phase I imaging results after synchronization, including the extracted target. ($\rm TO = 100~ns,~CD = 0.5621~m^2$)]{
    \includegraphics[width=0.9\textwidth]{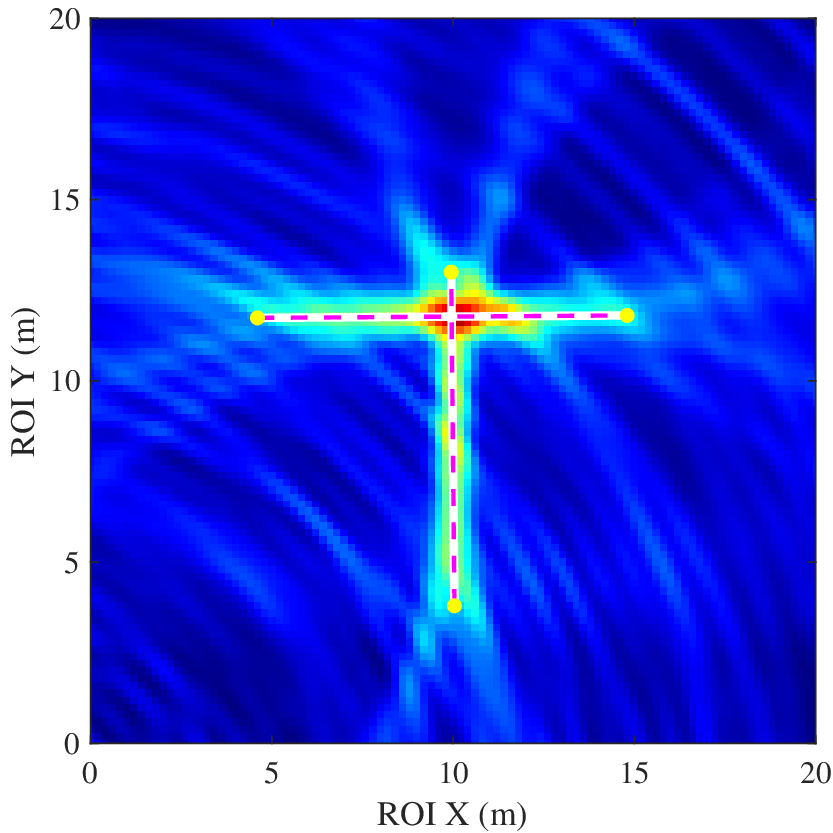}}
\end{minipage}
\begin{minipage}[t]{0.01\textwidth}\end{minipage}
\begin{minipage}[t]{0.22\textwidth}
    \centering
    \subfigure[Ideal Phase II imaging results based on GAMP, where the target location is perfectly on-grid. ($\rm RMSE = 0.0247~m$)]{
    \includegraphics[width=0.9\textwidth]{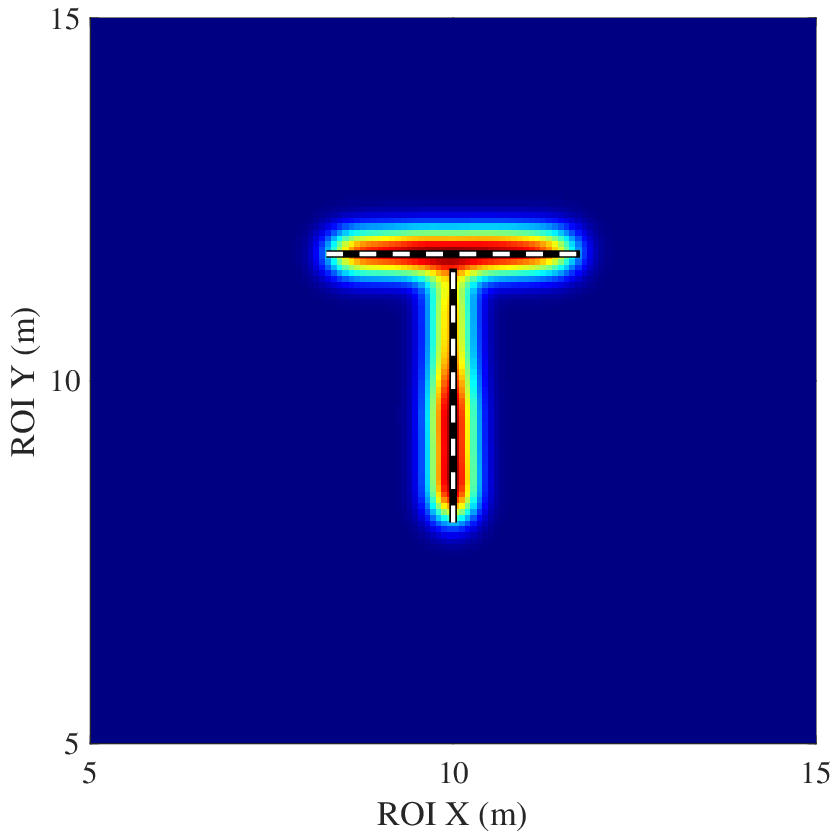}}
\end{minipage}

\begin{minipage}[t]{0.22\textwidth}
    \centering
    \subfigure[Phase II imaging results based on GAMP, where the target is an off-grid extended target. ($\rm RMSE = 0.4460~m$)]{
    \includegraphics[width=0.9\textwidth]{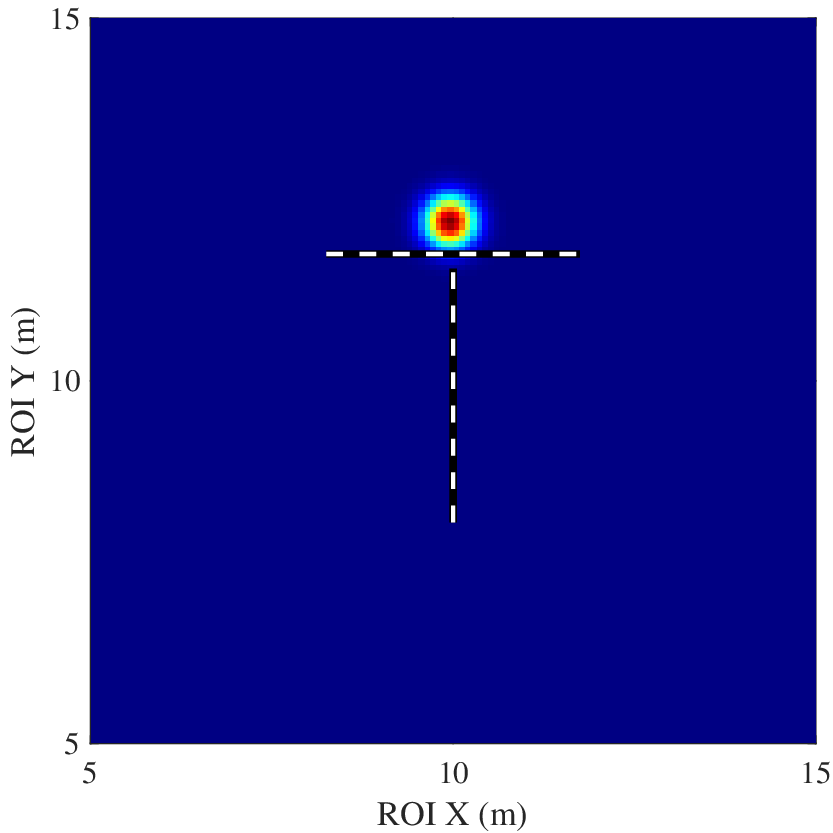}}
\end{minipage}
\begin{minipage}[t]{0.01\textwidth}\end{minipage}
\begin{minipage}[t]{0.22\textwidth}
    \centering
    \subfigure[Phase II imaging results based on OG-AMP, where the target is an off-grid extended target. ($\rm RMSE = 1.9057~m$)]{
    \includegraphics[width=0.9\textwidth]{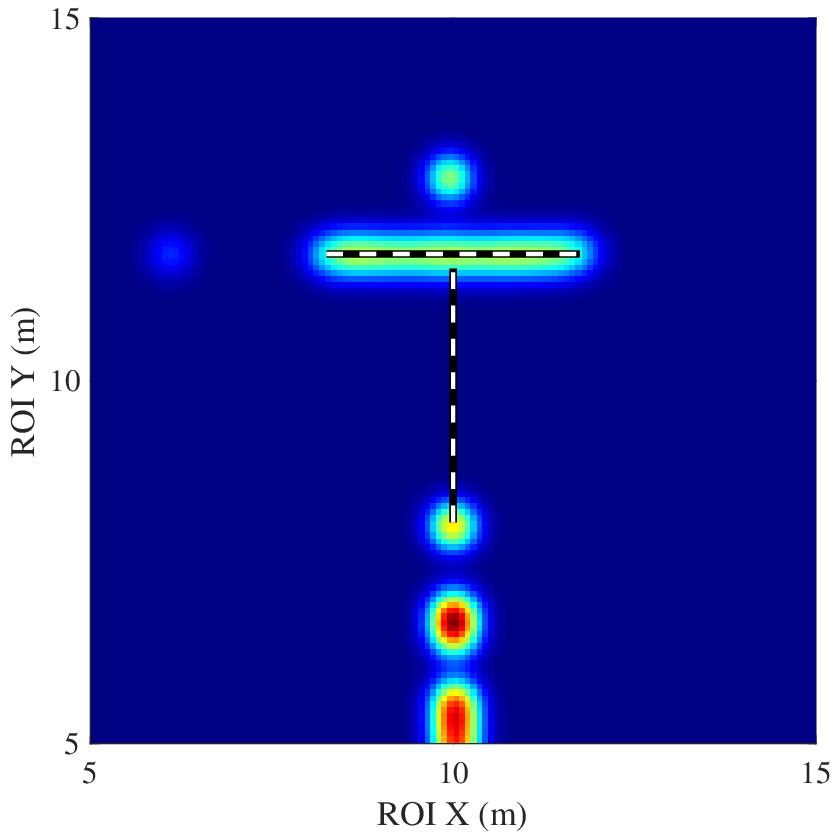}}
\end{minipage}
\begin{minipage}[t]{0.01\textwidth}\end{minipage}
\begin{minipage}[t]{0.22\textwidth}
    \centering
    \subfigure[Phase II BiOG-AMP imaging for an off-grid extended target without occlusion and dictionary matrix update. ($\rm RMSE = 1.3393~m$)]{
    \includegraphics[width=0.9\textwidth]{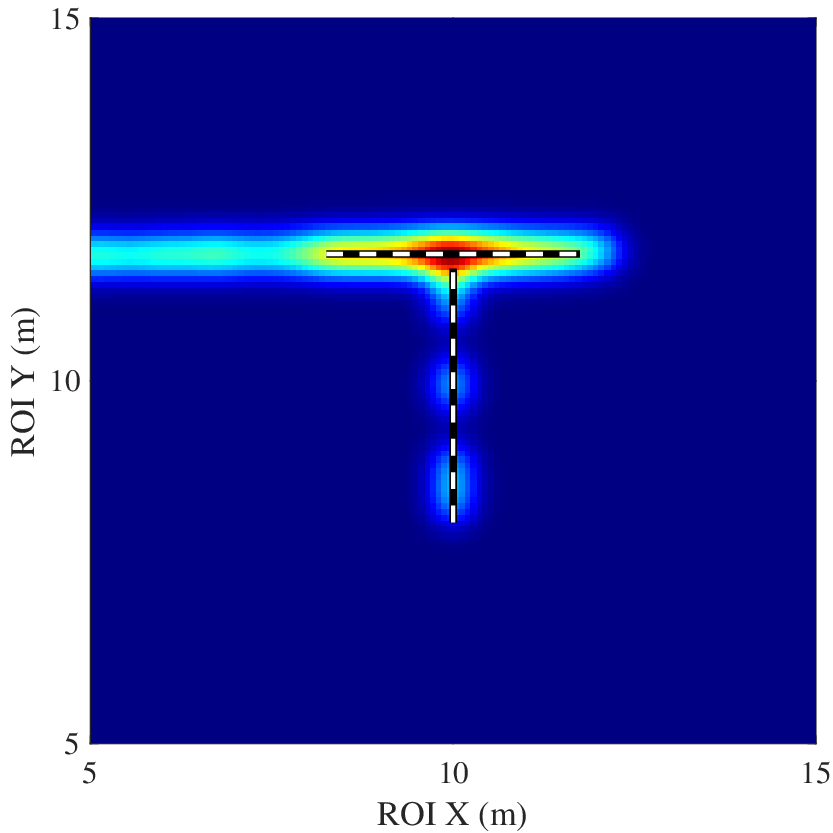}}
\end{minipage}
\begin{minipage}[t]{0.01\textwidth}\end{minipage}
\begin{minipage}[t]{0.22\textwidth}
    \centering
    \subfigure[Phase II imaging results based on complete BiOG-AMP, where the target is an off-grid extended target. ($\rm RMSE = 0.0227~m$)]{
    \includegraphics[width=0.9\textwidth]{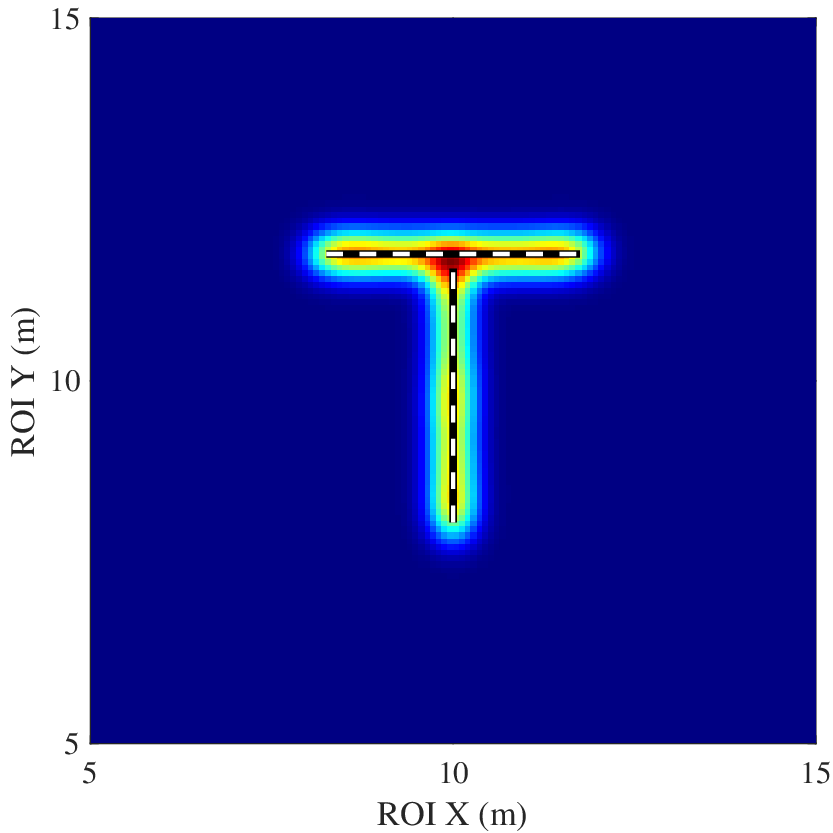}}
\end{minipage}
\caption{The imaging results of GAMP, OG-AMP and the proposed BiOG-AMP method. (a)-(c): Phase I performed effective synchronization and initialization. (d)-(e): The performance degradation of GAMP in Phase II by off-grid and orientation offsets. (f)-(h): The effectiveness of orientation shift and matrix update in the proposed BiOG-AMP.}
\label{fig:demo}
\vspace{-0.5cm}
\end{figure*}
\vspace{-0.3cm}
\subsection{Simulation Setting and Metrics}
Unless otherwise specified, the default simulation parameters are set as follows. We consider a \ac{D-ISAC} system with $N_{\rm Tx} = 30$ Txs and $N_{\rm Rx} = 30$ Rxs surrounding a $20\text{ m} \times 20\text{ m}$ \ac{2D} \ac{ROI}. Phase I uses a $N = 100 \times 100$ pixel grid, while Phase II applies a finer $0.1\text{ m} \times 0.1\text{ m}$ spacing around identified targets. The system operates at $f_{\rm c} = 3~\text{GHz}$ with $100~\rm{MHz}$ bandwidth, utilizing $128$ subcarriers in Phase I and $2$ in Phase II for computational efficiency as spatial resolution is aperture-dominated. The \ac{SNR} is $20~\text{dB}$ under complex AWGN with variance $\sigma_{\rm n}^2$. Synchronization impairments include initial \acp{TO} $\sim {\rm U}[0, 100~\text{ns}]$ and \acp{PO} $\sim {\rm U}[0, 2\pi)$. For the \ac{BiOG-AMP} algorithm, tolerance $\epsilon = 10^{-4}$, sparsity $\eta = 5\%$, and priors $\mu_{x,n} = 1, \sigma_{x,n}^2 = 0.1$, with initial offset variances derived from Phase I. Performance curves are averaged over $1000$ Monte Carlo runs.

To evaluate performance across different stages, two metrics are employed:
\begin{itemize}
    \item In Phase I, since the extracted targets are continuous line segments, \ac{CD} is used to evaluate the imaging accuracy. To compute the \ac{CD}, the ground-truth line segments and the extracted segments are uniformly discretized into fine point sets $\mathcal{S}$ and $\hat{\mathcal{S}}$, respectively. The \ac{CD} is defined as
    \begin{align}
    d_{\mathrm{CD}} = \frac{1}{|\mathcal{S}|} \sum_{\bm{p} \in \mathcal{S}} \min_{\hat{\bm{p}} \in \hat{\mathcal{S}}} \|\bm{p} - \hat{\bm{p}}\|_2^2 
    + \frac{1}{|\hat{\mathcal{S}}|} \sum_{\hat{\bm{p}} \in \hat{\mathcal{S}}} \min_{\bm{p} \in \mathcal{S}} \|\hat{\bm{p}} - \bm{p}\|_2^2,
    \end{align}
    where $\bm{p}$ and $\hat{\bm{p}}$ represent the 2D spatial coordinates of point samples in the respective sets. A smaller $d_{\mathrm{CD}}$ indicates higher spatial layout similarity. 

    \item In Phase II, \ac{RMSE} is adopted to measure the imaging accuracy of active target pixels relative to the true continuous target profiles. Let $\mathcal{A}$ denote the set of estimated active pixels with coordinates $\hat{\bm{p}}_n, n \in \mathcal{A}$. The RMSE is defined as the average Euclidean distance from each estimated active pixel location to the nearest point on the true target line segments $\mathcal{L}$,
    \begin{equation}
    \mathrm{RMSE} = \sqrt{ \frac{1}{|\mathcal{A}|} \sum_{n \in \mathcal{A}} \min_{\bm{p} \in \mathcal{L}} \|\hat{\bm{p}}_n - \bm{p}\|_2^2 }.
    \end{equation}
\end{itemize}

\begin{figure*}[t]
  \centering
  \begin{minipage}[t]{0.32\textwidth}
    \centering
    \includegraphics[width=0.9\textwidth]{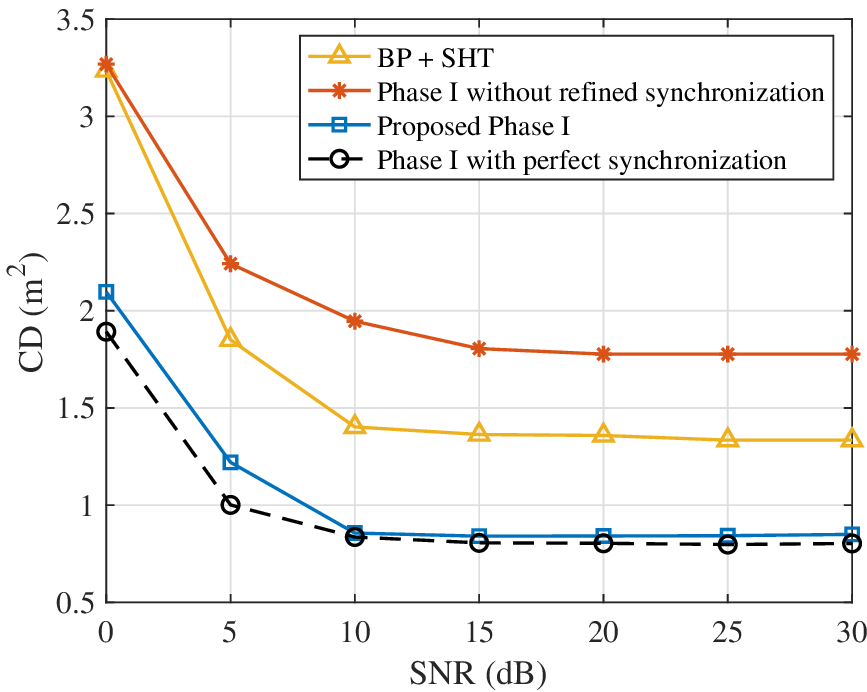}
    \caption{The relationship between the SNR and Phase I imaging performance.}
    \label{fig:SNR_P1}
  \end{minipage}
  \hfill 
  \begin{minipage}[t]{0.32\textwidth}
    \centering
    \includegraphics[width=0.9\textwidth]{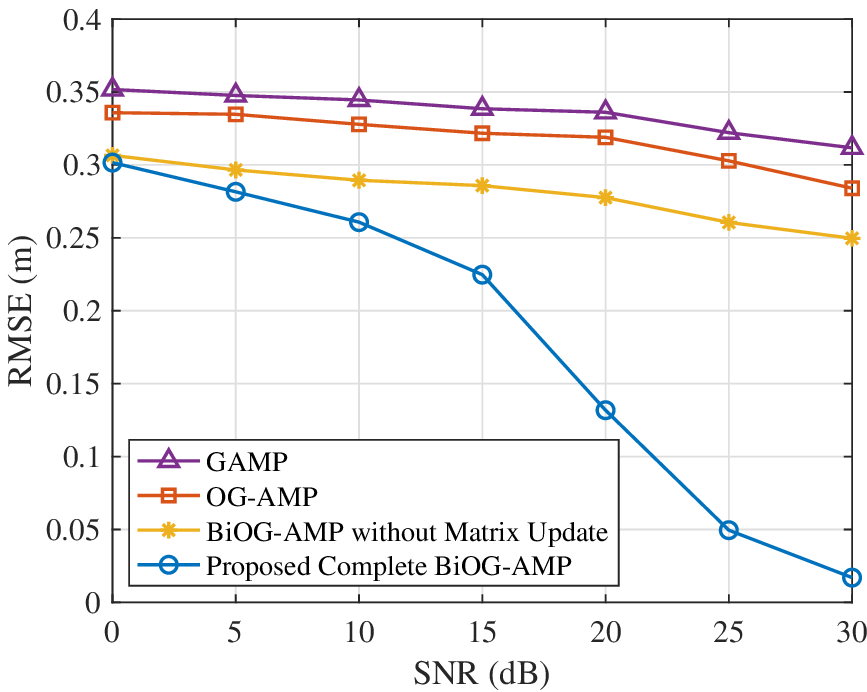}
    \caption{The relationship between the SNR and Phase II imaging performance.}
    \label{fig:SNR_P2}
  \end{minipage}
  \hfill
  \begin{minipage}[t]{0.32\textwidth}
    \centering
    \includegraphics[width=0.9\textwidth]{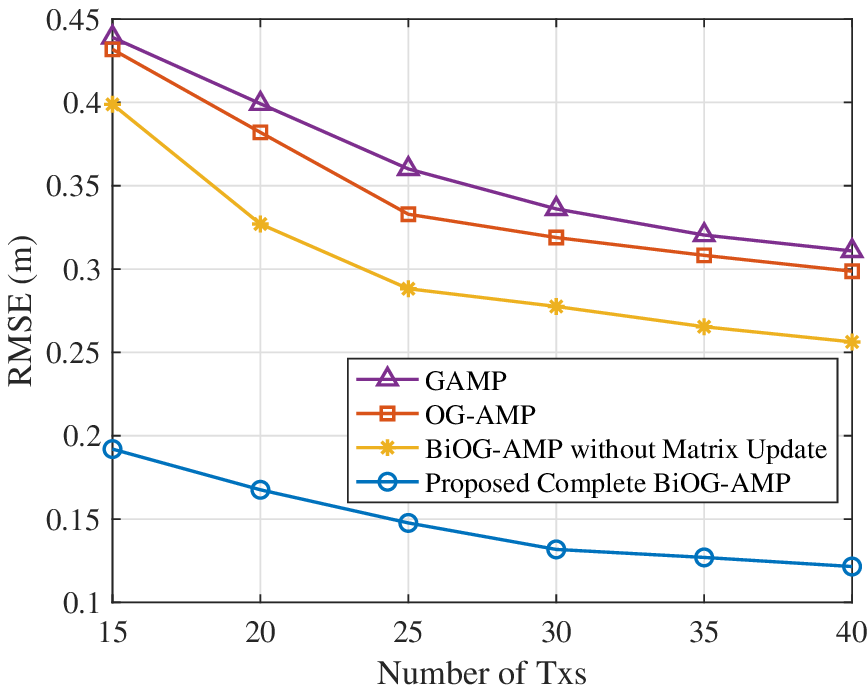}
    \caption{The relationship between the number of Txs and Phase II imaging performance.}
    \label{fig:Tx_P2}
  \end{minipage}
  \vspace{-0.5cm}
\end{figure*}
\begin{figure*}[t]
  \centering
  \begin{minipage}[t]{0.32\textwidth}
    \centering
    \includegraphics[width=0.9\textwidth]{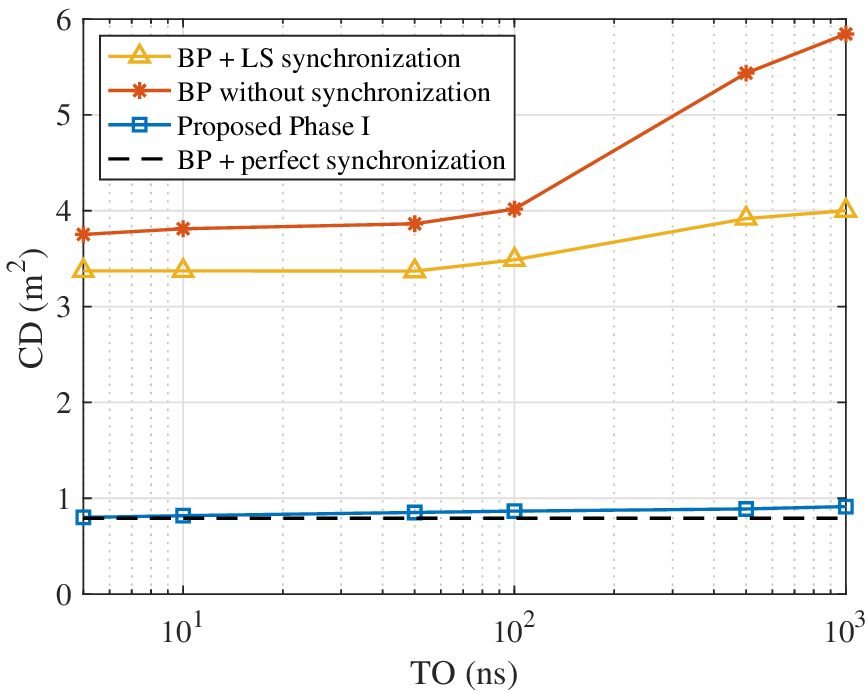}
    \caption{The relationship between the TO and Phase I imaging performance.}
    \label{fig:TO_P1}
  \end{minipage}
  \hfill 
  \begin{minipage}[t]{0.32\textwidth}
    \centering
    \includegraphics[width=0.9\textwidth]{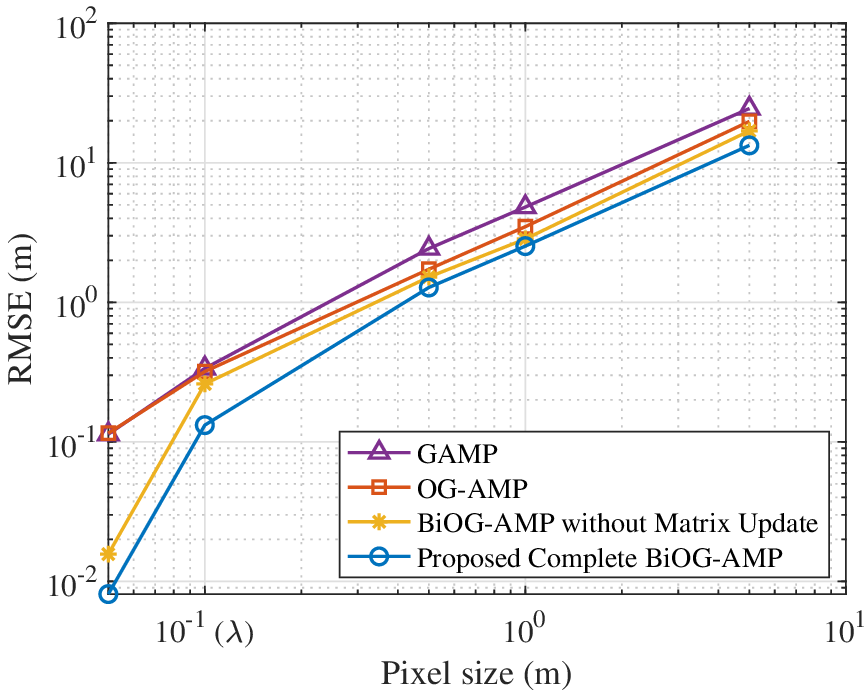}
    \caption{The relationship between the pixel size and Phase II imaging performance.}
    \label{fig:psize_P2}
  \end{minipage}
  \hfill
  \begin{minipage}[t]{0.32\textwidth}
    \centering
    \includegraphics[width=0.9\textwidth]{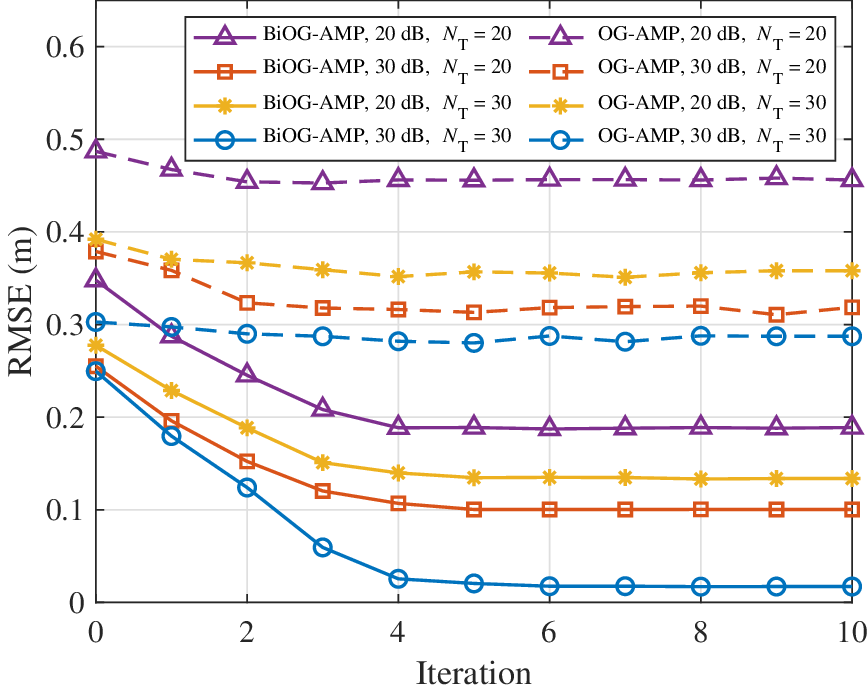}
    \caption{The convergence performance of Phase II.}
    \label{fig:conv_P2}
  \end{minipage}
  \vspace{-0.5cm}
\end{figure*}
\vspace{-0.3cm}
\subsection{Performance Evaluation}
Fig.~\ref{fig:demo} provides qualitative imaging comparisons between the baseline schemes and the proposed method on a representative T-shaped extended target. In Phase I (Fig.~\ref{fig:demo}(a)–(c)), the proposed incoherent \ac{BP} combined with \ac{PHT} successfully extracts the coarse geometric layout of the extended target even under severe synchronization errors.
In Phase II (Fig.~\ref{fig:demo}(d)–(h)), standard \ac{GAMP} and \ac{OG-AMP} fail to provide satisfactory reconstruction results when applied to continuous extended target profiles. As shown in Fig.~\ref{fig:demo}(e) and (f), spatial off-grid position offsets and surface orientation mismatches severely degrade reconstruction accuracy. Furthermore, neglecting the geometric occlusion effect leads to physically inconsistent reconstruction with spurious artifacts, as illustrated in Fig.~\ref{fig:demo}(g). In the contrast, the proposed complete \ac{BiOG-AMP} algorithm (Fig.~\ref{fig:demo}(h)) perfectly restores the complete profile.

Fig.~\ref{fig:SNR_P1} illustrates the Phase I performance versus \ac{SNR}. As analyzed in Section~\ref{sec:PhaseI}, suppressing strong direct \ac{LOS} paths and solving a graph-based linear system effectively maps link-level timing offsets to global node-level delays. Consequently, Phase I maintains low \ac{CD} values across all \ac{SNR} regimes, closely approaching the ideal performance and providing a reliable geometric prior for Phase II.

Fig.~\ref{fig:SNR_P2} depicts Phase II imaging \ac{RMSE} versus \ac{SNR} across baseline algorithms. Standard \ac{GAMP} plateaus near $0.35~\text{m}$ due to grid quantization mismatch. Although \ac{OG-AMP} compensates for spatial position offsets, it degrades under unmodeled surface orientation offsets and anisotropic scattering. In contrast, the proposed \ac{BiOG-AMP} outperforms all benchmarks, achieving millimeter-level precision at high \ac{SNR}.

Fig.~\ref{fig:Tx_P2} evaluates the impact of $N_{\rm Tx}$ on Phase II imaging performance, with $N_{\rm Rx} = 30$. Consistent with the analyses in Sections~\ref{sec:perA} and \ref{sec:perB}, increasing $N_{\rm Tx}$ enhances spatial multi-view diversity and accumulated \ac{FI}. While \ac{GAMP} and \ac{OG-AMP} suffer from model mismatches, \ac{BiOG-AMP} effectively leverages spatial diversity to steadily reduce reconstruction \ac{RMSE} as more transmitters are deployed.
\vspace{-0.3cm}
\subsection{Robustness and Convergence Evaluation}
Fig.~\ref{fig:TO_P1} evaluates Phase I robustness against varying initial \acp{TO}. Without synchronization, the \ac{BP} map degrades severely. Standard \ac{BP} with \ac{LS} treats links independently, failing to exploit global network constraints. In contrast, our approach maintains low \ac{CD} values across wide \acp{TO}, closely approaching ideal performance under perfect synchronization to provide a reliable prior for Phase II.

Fig.~\ref{fig:psize_P2} evaluates performance under varying pixel sizes, validating the discretization analysis in Section~\ref{sec:perC}. Larger pixel sizes scale up off-grid offsets, where the linearization error scales with $d_{\text{pixel}}^4$, worsening dictionary mismatch and degrading accuracy. Despite this inherent loss, the proposed BiOG-AMP algorithm consistently outperforms baseline methods.

Fig.~\ref{fig:conv_P2} illustrates algorithm convergence over an extended iteration range. \ac{BiOG-AMP} converges stably, achieving major gains within a few iterations. In contrast, \ac{OG-AMP} fails to converge effectively due to uncompensated spatial-orientation mismatches.
\vspace{-0.3cm}
\section{Conclusion}
This paper presents a hierarchical computational imaging framework for extended target reconstruction in \ac{D-ISAC} systems. Phase I eliminates timing and phase synchronization errors via LOS suppression, topological constraints, and probabilistic shape priors. Phase II incorporates anisotropic facet reflections and complex occlusions into an off-grid scattering model, where the proposed BiOG-AMP algorithm mitigates grid mismatches while enforcing physical consistency across transceivers. Derivations and simulations demonstrate high reconstruction accuracy and robustness against synchronization errors and occlusions, offering a practical solution for future ISAC sensing.

\begin{appendices}
\section{Derivation of Reflection Gradients}
\label{app:reflection_gradients}
When $\bm{n} \cdot (-\bm{l}_m(\rho)) > 0$, the diffuse reflection gradient is calculated as
    $\frac{\partial \alpha_{m}^{\rm diff}}{\partial \vartheta_n} = \frac{\sqrt{k_{\rm d}}}{2L_m^{\rm d}(\rho) \sqrt{\bm{n} \cdot -\bm{l}_m(\rho)}} \left( -\bm{u}_{{\rm Tx},m}(\rho) \cdot \bm{n}' \right)$,
and is defined as zero otherwise. Similarly, for $\bm{u}_{{\rm Rx},m}(\rho) \cdot \bm{r}_m(\rho) > 0$, the specular reflection gradient is calculated as
$\frac{\partial \alpha_{m}^{\rm spec}}{\partial \vartheta_n} = \frac{\sqrt{k_s} \gamma}{2L_m^{\rm s}(\rho)} \left( \bm{u}_{{\rm Rx},m}(\rho) \cdot \bm{r}_m(\rho) \right)^{\gamma/2 - 1} ( \bm{u}_{{\rm Rx},m}(\rho) \cdot \frac{\partial \bm{r}_m(\rho)}{\partial \vartheta_n} )$,
and is zero otherwise, where $\bm{l}_m(\rho)$ and $\bm{r}_m(\rho)$ represent the incident and perfect specular reflection vectors evaluated at coordinate $\bar{\bm{p}}_{n, \rho}$, and 
${\partial \bm{r}_m(\rho)}/{\partial \vartheta_n} = 2 \left[ (\bm{u}_{{\rm Tx},m}(\rho) \cdot \bm{n}')\bm{n} + (\bm{u}_{{\rm Tx},m}(\rho) \cdot \bm{n})\bm{n}' \right]$.

\section{Approximation of Message $\mu_{{\rm F}_{y,m,k}\rightarrow x_{n}}^{(t)}(x_n)$}\label{ap:A}
According to the factor graph depicted in Fig.~\ref{fig:fg}, the message from the factor node ${\rm F}_{y,m,k}$ to the variable $x_n$ is computed by marginalizing the joint likelihood over all other involved variables,
    $\mu^{(t)}_{{\rm F}_{y,m,k} \rightarrow x_n }(x_n) \propto  
    \int \Pr(y_{m,k}|\bm{z}) \prod_{j \neq n}\mu^{(t)}_{x_j \rightarrow {\rm F}_{y,m,k}}(x_j){\rm d}x_j
     \times 
    \prod_{l = 1}^N \Big( \prod_{i \in \{s_x, s_y, s_\vartheta\}} \mu^{(t)}_{z_{l,i} \rightarrow {\rm F}_{y,m,k}}(z_{l,i}){\rm d}z_{l,i} \Big)$.
By isolating the variable $x_n$, the observed signal $y_{m,k}$ is expressed as
    $y_{m,k} =  a_{m,k,n}^{(x)} x_n + \Big( \sum_{j \neq n} a_{m,k,j}^{(x)} x_j 
    + \sum_{l=1}^N \sum_{i \in \{s_x, s_y, s_\vartheta\}} a_{m,k,l}^{(i)} z_{l,i} + n_{m,k} \Big), 
    =  a_{m,k,n}^{(x)} x_n + S_{m,k \rightarrow x_n}$
Invoking the \ac{CLT} for a large number of scatterers ($N \rightarrow \infty$), in $t$-th iteration, $S_{m,k \rightarrow x_n}^{(t)}$ is approximated as a complex Gaussian random variable $\mathcal{CN}(S_{m,k \rightarrow x_n}^{(t)} ; Z_{m,k \rightarrow x_n}^{(t)}, V_{m,k \rightarrow x_n}^{(t)})$. 
The message integration then simplifies as
$\mu_{{\rm F}_{y,m,k}\rightarrow x_{n}}^{(t)}(x_n) \propto \int \mathcal{CN}(y_{m,k} ; a_{m,k,n}^{(x)} x_n + S_{m,k \rightarrow x_n}^{(t)}, \sigma_{{\rm w}}^2) 
 \times \mathcal{CN}(S_{m,k \rightarrow x_n}^{(t)} ; Z_{m,k \rightarrow x_n}^{(t)}, V_{m,k \rightarrow x_n}^{(t)}) {\rm d}S_{m,k \rightarrow x_n}^{(t)} 
\propto \mathcal{CN}(x_n; \hat x_{m,k \rightarrow x_n}^{(t)}, v_{m,k \rightarrow x_n}^{(t)})$.
where $\hat x_{m,k \rightarrow x_n}^{(t)}$ and $v_{m,k \rightarrow x_n}^{(t)}$ are the mean and variance of $\mu_{{\rm F}_{\hat y,m,k}\rightarrow x_{n}}^{(t)}(x_n)$ as calculated in \eqref{eq:z_update_mean} and \eqref{eq:z_update_var}. The mean $Z_{m,k \rightarrow z_{n,i}}^{(t)}$ and variance $V_{m,k \rightarrow z_{n,i}}^{(t)}$ of $S_{m,k \rightarrow x_n}^{(t)}$ are calculated by summing the contributions from the incoming messages $\mu_{z_{n,i} \to {{\rm F}_{y,m,k}}}^{(t)}$,
\begin{align}
Z_{m,k \rightarrow n,i}^{(t)} =\;& \sum_{j \neq n} a_{m,k,j}^{(x)} \hat{x}_{j \to {\rm F}_{y,m,k}}^{(t)} \nonumber\\ &+ \sum_{l=1}^N \sum_{i \in \{s_x, s_y, s_\vartheta\}} a_{m,k,l}^{(i)} \hat{z}_{l,i \to {\rm F}_{y,m,k}}^{(t)}, \label{eq:zmtxn}
\end{align}
\begin{align}
V_{m,k \rightarrow n,i}^{(t)} =\;& \sum_{j \neq n} |a_{m,k,j}^{(x)}|^2 v_{j \to {\rm F}_{y,m,k}}^{(t)} \nonumber \\ &+ \sum_{l=1}^N \sum_{i \in \{s_x, s_y, s_\vartheta\}} |a_{m,k,l}^{(i)}|^2 v_{{l,i} \to {\rm F}_{y,m,k}}^{(t)}, \label{eq:vvmtxn}
\end{align}
where $\hat{z}_{l,i \to {\rm F}_{y,m,k}}^{(t)}$ and $v_{{l,i} \to {\rm F}_{y,m,k}}^{(t)}$ are the mean and variance of the messages $\mu_{z_{n,i} \rightarrow {\rm F}_{y,m,k}}^{(t)}(z_{n,i})$ from \ac{VN} $\bm z_n$ to \ac{FN} ${\rm F}_{y}$, which will be derived later.

\section{Approximation of Message $\mu_{x_n \rightarrow {\rm F}_{z,n}}^{(t)}(x_n)$}\label{ap:B}
The message from \ac{VN} $x_n$ to \ac{FN} ${\rm F}_{z,n}$ is calculated as the product of all messages incoming from \acp{FN} ${\rm F}_{y,m,k}$,
$\mu_{x_n \rightarrow {\rm F}_{z,n}}^{(t)}(x_n) \propto \prod_{m=1}^{M} \prod_{k=1}^{K} \mu_{{\rm F}_{y,m,k} \rightarrow x_n}^{(t)}(x_n)$.
Substituting the Gaussian approximations derived in Appendix~\ref{ap:A}, we have
$\mu_{x_n \rightarrow {\rm F}{z,n}}^{(t)}(x_n) \propto \prod_{m=1}^{M} \prod_{k=1}^{K} \mathcal{CN}(x_n; \hat x_{{m,k} \rightarrow x_n}^{(t)}, v_{{m,k} \rightarrow x_n}^{(t)})
\propto \mathcal{CN}(x_n; \hat{x}_{n \rightarrow {\rm F}_z}^{(t)}, v_{n \rightarrow {\rm F}_z}^{(t)})$,
where $\hat{x}_{n \to {\rm F}_z}^{(t)}$ and $v_{n \to {\rm F}_z}^{(t)}$ are the mean and variance of $\mu_{x_{n} \rightarrow {\rm F}_{z,n}}^{(t)}(x_n)$, then we obtain \eqref{eq:z_vn_to_fn_mean} and \eqref{eq:z_vn_to_fn_var}.

\section{Derivation of $Z_{z_{n,i}}^{(t)}(c_n)$}\label{ap:C}
According to Appendix~\ref{ap:B}, the message $\mu_{z_{n,i}\rightarrow {\rm F}_{z,n}}^{(t)}(z_{n,i})$ is approximated as a Gaussian distribution. We consider the two possible states of $c_n$: 
\begin{itemize}
    \item When $c_n = 1$, the prior is $\Pr(z_{n,i}|c_n=1) = \mathcal{CN}(z_{n,i}; \mu_{z,n,i}, \sigma_{z,n,i}^2)$. The term $Z_{z_{n,i}}^{(t)}(1)$ is calculated as
    $Z_{z_{n,i}}^{(t)}(1) = \int \mathcal{CN}(z_{n,i}; \mu_{z,n,i}, \sigma_{z,n,i}^2)  \mathcal{CN}(z_{n,i}; \hat{z}_{n,i\rightarrow {\rm F}_z}^{(t)}, v_{n,i\rightarrow {\rm F}_z}^{(t)}) {\rm d}z_{n,i} 
    = \mathcal{CN}(\mu_{z,n,i} ; \hat{z}_{n,i\rightarrow {\rm F}_z}^{(t)}, \sigma_{z,n,i}^2 + v_{n,i\rightarrow {\rm F}_z}^{(t)})$.

    \item When $c_n = 0$, the prior is $\Pr(z_{n,i}|c_n=0) = \delta(z_{n,i})$. The term $Z_{z_{n,i}}^{(t)}(0)$ is calculated as
    $Z_{z_{n,i}}^{(t)}(0) = \int \delta(z_{n,i})\mathcal{CN}(z_{n,i}; \hat{z}_{n,i\rightarrow {\rm F}_z}^{(t)}, v_{n,i\rightarrow {\rm F}_z}^{(t)}) {\rm d}z_{n,i} 
    = \mathcal{CN}(0 ; \hat{z}_{n,i\rightarrow {\rm F}_z}^{(t)}, v_{n,i\rightarrow {\rm F}_z}^{(t)})$.
\end{itemize}

\section{Approximation of Message $\mu_{{\rm F}_{z,n}\rightarrow x_{n}}^{(t+1)}(x_{n})$}\label{ap:D}
The message from the prior factor node ${\rm F}_{z,n}$ to the variable $x_n$ is obtained by summing over the states of the sparsity variable $c_n$ and integrating over the offset variables
\begin{align} 
\mu^{(t+1)}_{{\rm F}_{z,n} \rightarrow x_{n}}(x_n) = &\sum_{c_n \in \{0,1\}} \Pr(x_n | c_n) \mu_{c_n \to {\rm F}{z,n}}(c_n) \nonumber \\ & \prod_{j \in \{s_x, s_y, s_\vartheta\}} Z_{z_{n,j}}^{(t)}(c_n), \label{eq:mfxntxn_appendix}
\end{align}
We evaluate the two cases for $c_n$:
\begin{itemize}
\item When $c_n = 1$, the corresponding term in the \eqref{eq:mfxntxn_appendix} is
${\rm Term}1 = \mathcal{CN}(x_{n};\mu_{x,n},\sigma_{x,n}^{2})\eta \prod_{j \in \{s_x, s_y, s_\vartheta\}} Z_{z_{n,j}}^{(t)}(1)$.
\item When $c_n = 0$, the corresponding term in the \eqref{eq:mfxntxn_appendix} is
${\rm Term}0 = \delta(x_n) (1-\eta) \prod_{j \in \{s_x, s_y, s_\vartheta\}} Z_{z_{n,j}}^{(t)}(0)$.
\end{itemize}
By defining the normalized weights $k_{n,0}^{(t+1)}$ and $k_{n,1}^{(t+1)}$ as in \eqref{eq:kn0_final} and \eqref{eq:kn1_final}, the message is expressed as the weighted sum of these two terms, we obtain \eqref{eq:fzntxn_final}.

\section{Approximation of Message $\mu_{z_{n,i} \rightarrow {\rm F}_{y,m,k}}^{(t+1)}(z_{n,i})$}\label{ap:E}
The message $\mu_{z_{n,i} \rightarrow {\rm F}_{y,m,k}}^{(t+1)}(z_{n,i})$ at the $(t+1)$-th iteration is the product of the incoming message from the prior node and all observation nodes except the current $(m,k)$,
\begin{align}
\mu_{z_{n,i} \rightarrow {\rm F}{y,m,k}}^{(t+1)}(z_{n,i}) \propto &\;\mu_{{\rm F}_{z,n} \rightarrow z_{n,i}}^{(t+1)}(z_{n,i}) \nonumber\\ & \prod_{(m',k') \ne (m,k)} \mu_{{\rm F}_{y,m',k'} \rightarrow z_{n,i}}^{(t)}(z_{n,i}).
\label{eq:cavity_product}
\end{align}
The second term in \eqref{eq:cavity_product} represents the cavity message, which is Gaussian,
\begin{align}
\prod_{(m',k') \ne (m,k)} & \mu_{{\rm F}{y,m',k'} \to z_{n,i}}^{(t)}(z_{n,i}) \\
&\propto \frac{\mathcal{CN}(z_{n,i};\hat{z}_{n,i\rightarrow {\rm F}_{z}}^{(t)},v_{n,i\rightarrow {\rm F}_{z}}^{(t)})}{\mathcal{CN}(z_{n,i};\hat{z}_{m,k \rightarrow z_{n,i}}^{(t)},v_{m,k \rightarrow z_{n,i}}^{(t)})} \nonumber \\
&\propto \mathcal{CN}(z_{n,i}; \hat{z}_{{\rm cav}, n,i, m,k}^{(t)}, v_{{\rm cav}, n,i, m,k}^{(t)}).
\end{align}
The cavity variance and mean are calculated as
\begin{equation}
v_{{\rm cav}, n,i, m,k}^{(t)} = \Big(\frac{1}{v_{n,i\rightarrow {\rm F}_{z}}^{(t)}} - \frac{1}{v_{m,k \rightarrow z_{n,i}}^{(t)}}\Big)^{-1}, \label{eq:cavv}
\end{equation}
\begin{equation}
\hat{z}_{{\rm cav}, n,i, m,k}^{(t)} = v_{{\rm cav}, n,i, m,k}^{(t)} \Big( \frac{\hat{z}_{n,i\rightarrow {\rm F}_{z}}^{(t)}}{v_{n,i\rightarrow {\rm F}_{z}}^{(t)}} - \frac{\hat{z}_{m,k \rightarrow z_{n,i}}^{(t)}}{v_{m,k \rightarrow z_{n,i}}^{(t)}} \Big). \label{eq:cavz}
\end{equation}
Substituting the Bernoulli-Gaussian prior from \eqref{eq:fzntxn_final} and the cavity Gaussian into \eqref{eq:cavity_product}, we obtain:
\begin{align}
& \mu_{z_{n,i} \rightarrow {\rm F}_{y,m,k}}^{(t+1)}(z_{n,i}) \propto W_{n,i,0}^{(t+1)} \delta(z_{n,i}) \nonumber \\& + W_{n,i,1}^{(t+1)} \mathcal{CN}(z_{n,i}; \hat{z}_{{\rm prod}, n,i, m,k}^{(t+1)}, v_{{\rm prod}, n,i, m,k}^{(t+1)}),
\end{align}
where the weights are calculated as
\begin{align}
W_{n,i,0}^{(t+1)} &\propto k_{n,0}^{(t+1)} \mathcal{CN}(0; \hat{z}_{{\rm cav}, n,i, m,k}^{(t)}, v_{{\rm cav}, n,i, m,k}^{(t)}), \label{eq:Wni0}\\
W_{n,i,1}^{(t+1)} &\propto k_{n,1}^{(t+1)} \nonumber \\ & \mathcal{CN}(\mu_{z,n,i}; \hat{z}_{{\rm cav}, n,i, m,k}^{(t)}, \sigma_{z,n,i}^2 + v_{{\rm cav}, n,i, m,k}^{(t)}).\label{eq:Wni1}
\end{align}
The product distribution parameters are given by
\begin{align}
v_{{\rm prod}, n,i, m,k}^{(t+1)} &= \Big(\frac{1}{\sigma_{z,n,i}^2} + \frac{1}{v_{{\rm cav}, n,i, m,k}^{(t)}}\Big)^{-1}, \label{eq:prodv}\\
\hat{z}_{{\rm prod}, n,i, m,k}^{(t+1)} &= v_{{\rm prod}, n,i, m,k}^{(t+1)} \Big( \frac{\mu_{z,n,i}}{\sigma_{z,n,i}^2} + \frac{\hat{z}_{{\rm cav}, n,i, m,k}^{(t)}}{v_{{\rm cav}, n,i, m,k}^{(t)}} \Big).\label{eq:prodz}
\end{align}
Finally, the normalized activation probability is $P_{n,i,1 \to m,k}^{(t+1)} = W_{n,i,1}^{(t+1)}/({W_{n,i,0}^{(t+1)} + W_{n,i,1}^{(t+1)}})$. The mean $\hat{z}_{n,i \to m,k}^{(t+1)}$ and variance $v_{n,i \to m,k}^{(t+1)}$ in \eqref{eq:zntfymk_mean} and \eqref{eq:zntfymk_var} are then derived as the first and second central moments of this Bernoulli-Gaussian message.

\end{appendices}

\ifCLASSOPTIONcaptionsoff
  \newpage
\fi

\bibliographystyle{IEEEbib}
\bibliography{IEEEabrv, ref}
\end{document}

%% file: acronyms.tex
\acrodef{MMSE}{Minimum Mean Squared Error}

\acrodef{MSE}{mean square error}

\acrodef{PSD}{power spectral density}

\acrodef{RMSE}{root mean squared error}
\acrodef{SLR}{statistical linear regression}

\acrodef{AP}[AP]{access point}

\acrodef{ML}[ML]{maximum likelihood}
\acrodef{DBSCAN}[DBSCAN]{density-based spatial clustering of applications with noise}
\acrodef{UE}[UE]{user equipment}
\acrodef{BS}[BS]{base station}
\acrodef{VA}[VA]{virtual anchor}
\acrodef{VUE}[VUE]{virtual user equipment}
\acrodef{SP}[SP]{scattering  point}
\acrodef{IP}[IP]{incidence point}
\acrodef{fov}[FoV]{field-of-view}   
\acrodef{LOS}[LOS]{line-of-sight}
\acrodef{NLOS}[NLOS]{non-line-of-sight}
\acrodef{PMBM}[PMBM]{Poisson  multi-Bernoulli  mixture}
\acrodef{PMB(M)}[PMB(M)]{Poisson  multi-Bernoulli  (mixture)}
\acrodef{PMB}[PMB]{Poisson  multi-Bernoulli}
\acrodef{RFS}[RFS]{random finite set}
\acrodef{PPP}[PPP]{Poisson point process}
\acrodef{MBM}[MBM]{multi-Bernoulli  mixture}
\acrodef{MB}[MB]{multi-Bernoulli}

\acrodef{ekf}[EKF]{extended Kalman filter}
\acrodef{PDF}[PDF]{probability density function}

\acrodef{ckf}[CKF]{cubature Kalman filter}
\acrodef{rbp}[RBP]{Rao-Blackwellized particle}
\acrodef{gospa}[GOSPA]{generalized optimal subpattern assignment}
\acrodef{SLAM}[SLAM]{simultaneous localization and mapping}

\acrodef{TOA}[TOA]{time of arrival}
\acrodef{AOA}[AOA]{angles of arrival}
\acrodef{AOD}[AOD]{angles of departure}

\acrodef{IF}[IF]{information filter}
\acrodef{EIF}[EIF]{extended information filter}
\acrodef{kf}[KF]{Kalman filter}
\acrodef{PMF}[PMF]{probability mass function}
\acrodef{MAP}[MAP]{maximum a posteriori estimation}

\acrodef{DA}[DA]{data association}
\acrodef{OFDM}[OFDM]{orthogonal frequency division multiplexing}

\acrodef{PCRB}[PCRB]{posterior Cram{\'e}r-Rao bound}

\acrodef{EM}[EM]{expectation-maximization}
\acrodef{FOV}[FOV]{field of view}
\acrodef{RB}[RB]{Rao-Blackwellized }

\acrodef{ISAC}[ISAC]{integrated sensing and communication}

\acrodef{EK}[EK]{extended Kalman}
\acrodef{EKF}[EKF]{extended Kalman filter}

\acrodef{LMB}[LMB]{labeled multi-Bernoulli}
\acrodef{GLMB}[$\delta$-GLMB]{$\delta$-generalized labeled multi-Bernoulli}
\acrodef{PHD}[PHD]{probability hypothesis density}

\acrodef{OID}[OID]{optimal importance density}

\acrodef{MTT}[MTT]{multi-target tracking}

\acrodef{MCMC}[MCMC]{Markov chain Monte Carlo}

\acrodef{MH}[MH]{Metropolis-Hastings}

\acrodef{NMI}[NMI]{normalized mutual information}

\acrodef{CRB}[CRB]{Cram{\'e}r-Rao bound}

\acrodef{PIM}[PIM]{posterior information matrix}

\acrodef{OID}[OID]{optimal importance density}

\acrodef{MAE}[MAE]{mean absoluate error} 

\acrodef{FIM}[FIM]{Fisher information matrix}
\acrodef{FI}{Fisher information}
\acrodef{PIM}[PIM]{posterior information matrix}
\acrodef{PEB}[PEB]{position error bound}
\acrodef{LEB}[LEB]{landmark error bound}
\acrodef{HEB}[HEB]{heading error bound}
\acrodef{CEB}[CEB]{clock bias error bound}

\acrodef{AoA}[AoA]{angles of arrival}
\acrodef{AoD}[AoD]{angles of departure}

\acrodef{URA}[URA]{uniform rectangular array}

\acrodef{RTT}{round-trip-time}

\acrodef{LS}{least-squares}

\acrodef{CDF}{cumulative distribution function}

\acrodef{AWGN}{additive white Gaussian noise}

\acrodef{MUSIC}{multiple signal classification}

\acrodef{S-MUSIC}{sequential multiple signal classification}
\acrodef{OMP}{orthogonal matching pursuit}
\acrodef{SCM}{spatial covariance matrix}

\acrodef{MF}{matched filtering}

\acrodef{AIC}{akaike information criterion}
\acrodef{MDL}{minimum description length}
\acrodef{Tx}{transmitter}
\acrodef{Rx}{receiver}
\acrodef{CP}{cyclic prefix}
\acrodef{CSI}{channel state information}

\acrodef{LO}{local oscillator}
\acrodef{TO}{time offset}
\acrodef{PO}{phase offset}

\acrodef{IFFT}{inverse fast Fourier transform}
\acrodef{PDP}{power delay profile}
\acrodef{ROI}{region of interest}

\acrodef{BP}{back-projection}
\acrodef{PHT}{probabilistic Hough transform}
\acrodef{SHT}{standard Hough transform}
\acrodef{PCA}{principal component analysis}
\acrodef{SNR}{signal-to-noise ratio}
\acrodef{AMP}{approximate message passing}

\acrodef{OG-AMP}{off-grid approximate message passing}
\acrodef{BiOG-AMP}{bilinear off-grid approximate message passing}
\acrodef{MIMO}{multiple-input multiple-output}
\acrodef{CLT}{central limit theorem}

\acrodef{VN}{variable node}
\acrodef{FN}{factor node}
\acrodef{CPU}{central processing unit}
\acrodef{OTA}{over-the-air}
\acrodef{i.i.d.}{independent and identically distributed}

\acrodef{RIP}{restricted isometry property}
\acrodef{GAMP}{generalized approximate message passing}

\acrodef{CD}{chamfer distance}
\acrodef{2D}{2-dimensional}

\acrodef{RF}{radio frequency }
\acrodef{D-ISAC}{distributed ISAC}